\documentclass[journal]{IEEEtran}

\ifCLASSINFOpdf
\else
\fi
\usepackage{graphicx}
\usepackage{amsmath}
\usepackage{amssymb}
\usepackage{epstopdf}
\usepackage{subcaption}
\usepackage{xcolor}
\usepackage{hyperref}
\usepackage{fancyhdr}
\usepackage{algorithm,algpseudocode}

\fancypagestyle{copyright}{%
	\fancyhf{} % clear all header and footer fields
	\fancyhead[C]{\footnotesize { This article has been accepted for publication in IEEE Transactions on Mobile Computing. This is the author's version which has not been fully edited and content may change prior to final publication. Citation information: DOI 10.1109/TMC.2025.3581561}}
	
	\fancyfoot[C]{\parbox{\textwidth}{\footnotesize \textcopyright 2025 IEEE. All rights reserved, including rights for text and data mining and training of artificial intelligence and similar technologies. Personal use is permitted, but republication/redistribution requires IEEE permission. See https://www.ieee.org/publications/rights/index.html for more information.}}%
}

\begin{document}
%

\title{LiSec-RTF: Reinforcing RPL Resilience Against Routing Table Falsification Attack in 6LoWPAN}      

\author{Shefali~Goel, Vinod~Kumar~Jain,~\IEEEmembership{Senior Member, IEEE}, and~Abhishek Verma*\thanks{*~Corresponding author}% <-this % stops a space
% <-this % stops a space

\thanks{Shefali Goel is with Computer Science \& Engineering Discipline, PDPM Indian Institute of Information Technology, Design and Manufacturing Jabalpur, Madhya Pradesh, India (e-mail: 21pcso08@iiitdmj.ac.in).}% <-this % stops a space
\thanks{Vinod Kumar Jain is with Department of Computer Science and Engineering, ABV-Indian Institute of Information Technology and Management Gwalior, Morena Link Road, Gwalior, Madhya Pradesh, India, 474015 (e-mail: vkjain@iiitdmj.ac.in).}% <-this % stops a space
\thanks{Abhishek Verma is with Department of Information Technology, Babasaheb Bhimrao Ambedkar University, Lucknow 226025, Uttar Pradesh, India (e-mail: abhiverma866@gmail.com).}}

\markboth{Journal of \LaTeX\ Class Files,~Vol.~14, No.~8, August~2015}%
{Shell \MakeLowercase{\textit{et al.}}: Bare Demo of IEEEtran.cls for IEEE Journals}



\maketitle

\thispagestyle{copyright}

\begin{abstract}

\textcolor{black}{Routing Protocol for Low-Power and Lossy Networks (RPL) is an energy-efficient routing solution for IPv6 over Low-Power Wireless Personal Area Networks (6LoWPAN), recommended for resource-constrained devices. While RPL offers significant benefits, its security vulnerabilities pose challenges, particularly due to unauthenticated control messages used to establish and maintain routing information. These messages are susceptible to manipulation, enabling malicious nodes to inject false routing data. A notable security concern is the Routing Table Falsification (RTF) attack, where attackers forge Destination Advertisement Object (DAO) messages to promote fake routes via a parent node’s routing table. Experimental results indicate that RTF attacks significantly reduce packet delivery ratio, increase end-to-end delay, and leverage power consumption. Currently, no effective countermeasures exist in the literature, reinforcing the need for a security solution to prevent network disruption and protect user applications. This paper introduces a \underline{Li}ghtweight \underline{Sec}urity Solution against \underline{R}outing \underline{T}able \underline{F}alsification Attack (LiSec-RTF), leveraging Physical Unclonable Functions (PUFs) to generate unique authentication codes, termed ``Licenses.” LiSec-RTF mitigates RTF attack impact while considering the resource limitations of 6LoWPAN devices in both static and mobile scenarios. Our testbed experiments indicate that LiSec-RTF significantly improves network performance compared to standard RPL under RTF attacks, thereby ensuring reliable and efficient operation.}

\end{abstract}


% \fi
\IEEEpeerreviewmaketitle

\section{Introduction}

\IEEEPARstart{T}he Internet of Things (IoT) is a fast-growing technology that comprises several physical devices, sensors, and software for exchanging data across networks, spanning from local area networks to the broader Internet  \cite{grammatikis2019securing}. This connectivity empowers devices to communicate and share information among themselves. According to the McKinsey Global Institute report, IoT's economic influence is projected to range between $3.9$ to $11.1$ trillion by the end of the year $2025$ \cite{espinoza2020estimating}. One of the primary benefits of IoT is its ability to facilitate communication among devices that are situated at far distant locations by employing IPv6 addressing. Nevertheless, when deploying IoT in industrial sectors (IIoT) like industrial automation and advanced metering infrastructure (AMI) there is a specific need for an infrastructure that requires small power yet supports a longer lifetime while supporting IPv6 capabilities \cite{9148607, civerchia2017industrial}. This requirement is satisfied through the implementation of IPv6 over Low-Power Wireless Personal Area Networks (6LoWPAN). The devices operating within 6LoWPAN are characterized by their limited computational ability, limited storage, memory, and energy-efficient attributes \cite{8270652,mayzaud2016taxonomy}.
The key benefit of these resource-constrained devices is their ability to run exceptionally low-voltage and consume minimal energy. This enables them to operate for extended periods, often spanning several years. A diverse array of such devices, commonly called ultra-low-powered micro-controllers, are easily available in the market. 
% The applications of 6LoWPAN devices are present in various real-world scenarios, like smart grid systems, connected homes, healthcare solutions, smart transportation systems, etc.
\newline
Routing is an integral component of 6LoWPAN networks, facilitating communication between devices situated at a distance from each other \cite{8876683}. The conventional routing protocols such as Adhoc On-Demand Distance Vector (AODV), Dynamic Source Routing (DSR), and Open Shortest Path First (OSPF) are well-suited for Wireless Sensor Networks due to the resource-rich nature of these networks \cite{SHARMA202312}. However, in the case of 6LoWPAN, embedded devices are inherently resource-constrained. Consequently, these traditional routing protocols are not recommended for 6LoWPAN. Therefore, there is a need for an energy-efficient routing protocol to conserve the resources of 6LoWPAN. Addressing this challenge is quite complex in 6LoWPAN networks \cite{airehrour2016secure, al2023systematic}. To meet this requirement, the Internet Engineering Task Force (IETF) introduced the Routing Protocol for Low Power and Lossy Networks (RPL) 
\cite{winter2012rpl}. RPL is explicitly designed for IPv6-based Low-Power and Lossy Networks (LLNs), which include 6LoWPAN. It functions as a proactive protocol, establishing and maintaining a routing topology in advance to enable efficient and reliable communication among nodes in the network.
It is important to emphasize that RPL is still in its developmental phase \cite{winter2012rpl}, and its specifications are presented in RFC 6550. RPL demonstrates diverse attributes, including the capacity to modify control packet frequencies, dynamically regulate control packet transmission rates via the trickle algorithm, and compute routing metrics utilizing an Objective Function to accommodate multi-path topologies \cite{musaddiq2020routing, gaddour2012rpl, vasseur2011rpl, lamaazi2020comprehensive}. These characteristics of RPL make it well-suited for deployment in 6LoWPAN \cite{muzammal2020comprehensive}. Nevertheless, it is important to remember that RPL and IIoT are susceptible to many types of attacks that target the privacy and security of users \cite{ni2024machine, zilberman2024sprinkler}. This vulnerability arises from the fact that RPL employs a shared key for security, making it susceptible to compromise if an attacker gains access to the key \cite{butun2019security}. Furthermore, RPL does not provide packet confidentiality, which means that an attacker with the capability to eavesdrop on communication can potentially obtain sensitive information \cite{bang2022embof, omar2023comprehensive}.

A common cyber threat in 6LoWPAN networks is insider or outsider routing attacks. These attacks exploit vulnerabilities in RPL to target legitimate nodes, potentially causing substantial disruptions to the overall performance of the network \cite{verma2020security}.
\textcolor{black}{The objective of this paper is to explore the Routing Table Falsification Attack (RTF). This attack can occur when malicious nodes tamper with Destination Advertisement Object (DAO) control messages or create forged DAO to establish fictitious downward routes. This type of attack is feasible only when the network enables the storing mode \cite{albinali2025replay, krari2024rpl}.} The consequence of such an attack can include longer paths, increased network delays, higher packet drop rates, and even network congestion. It is essential to consider that this specific vulnerability and its impact on RPL networks have not been extensively studied or documented \cite {mayzaud2016taxonomy, DBLP:journals/csur/BangRKC23} and does not have any defense solution to address RTF attack \cite{albinali2024towards}.
This paper shows that the RTF attack decreases the packet delivery ratio, and increases the packet delay and power consumption. To counteract Routing Table Falsification (RTF) in RPL, we have introduced a secure variant of RPL known as LiSec-RTF. \textcolor{black}{While RPL is susceptible to a wide range of attacks,  the proposed LiSec-RTF framework specifically focuses on mitigating the RTF attack (an Identity spoofing attack). 
This attack involves the injection of malicious routing information, enabling an attacker to manipulate the network's routing tables. LiSec-RTF aims to detect and prevent such falsified updates by integrating lightweight trust-based validation mechanisms suited for resource-constrained IoT environments. 
In the proposed solution (LiSec-RTF), we introduce a modified DAO message, referred to as $DAO_{modified}$, which incorporates an authentication mechanism for verifying sensor nodes at the 6LoWPAN Border Router (6LBR). The Reserved field of the DAO packet is utilized to encapsulate a License, which consists of a unique bit sequence generated by performing an exclusive OR (XOR) operation on the node’s Physical Unclonable Function (PUF) data. When a child node unicasts a $DAO_{modified}$ message to the 6LBR, the router extracts the License from the Reserved field and verifies it against the License generated during the node’s registration phase. This process enables the 6LBR to authenticate whether the License is genuine i.e., derived from the registered PUF data. Very few changes have been made in the standard RPL implementation to incorporate LiSec-RTF. We have only updated the existing DAO processing mechanism of standard RPL implementation to perform authentication of source DAO’s.  
}

Some of the benefits of the proposed LiSec-RTF approach include: (1) accurate detection of attack; (2) mitigation of attack which improves the network's performance in static and mobile network environments; (3) the proposed solution does not introduce any memory overhead on the resource-constrained nodes. \textcolor{black}{The novelty of the proposed LiSec-RTF framework lies in its specific focus on detecting and mitigating RTF attack in RPL-based networks. In the literature, there is no specific defense mechanism exists that focused on mitigating RTF attack in RPL-based IoT networks. Although several IDS can detect and mitigate a range of RPL-based attacks, they are not effective against RTF attack due to fundamental differences in attack characteristics. RTF attack exhibit distinct behavioral patterns compared to attacks such as Forwarding Misbehavior, DAO Inconsistency, Hatchetman, DIO Suppression, Energy Depletion, Spam DIS, and Advanced Vampire attacks. The proposed LiSec-RTF framework specifically focuses on mitigating the RTF attack. This attack involves the injection of malicious routing information, enabling an attacker to manipulate the network's routing tables. LiSec-RTF aims to detect and prevent such falsified updates by integrating lightweight trust-based validation mechanisms suited for resource-constrained IoT environments. 
 }
The key points of the contribution are outlined below:

\begin{figure}
    \centering
    \includegraphics[width=.4\textwidth]{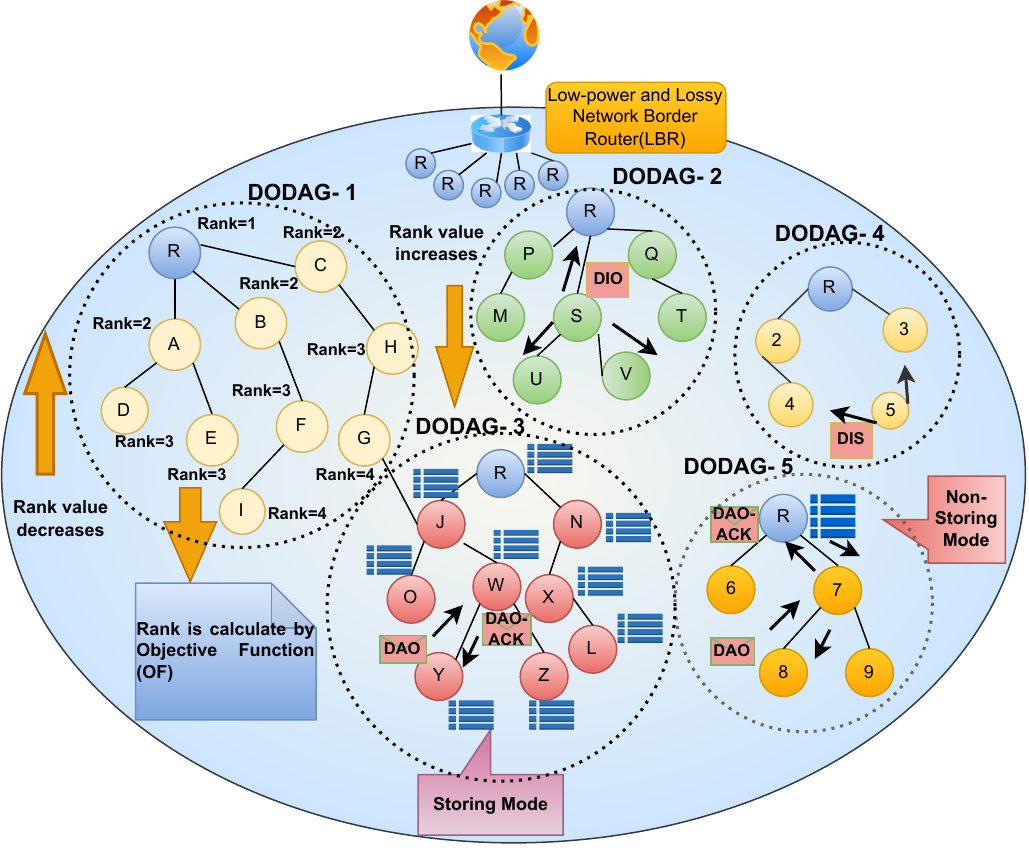}
    \caption{Overview of RPL}
    \label{fig:RPL}
\end{figure}
\begin{itemize}

 \item Analyze the impact of Routing Table Falsification attack on both static and mobile environment.
    \item An effective solution named ``LiSec-RTF" is proposed to mitigate the effect of Routing Table Falsification attack.
    \item The effectiveness of ``LiSec-RTF" is compared against the standard RPL protocol using simulation under Routing Table Falsification attack.
    \item A simulation study is performed comparing ``LiSec-RTF" against the standard RPL protocol in delivering good network performance while under Routing Table Falsification attack on both static and mobile scenarios.
     \item A testbed experimentation is also performed to validate the reported simulation results.
\end{itemize}

% \subsection{Organization of the paper}
The organization of the paper is as follows: Section \ref{Sec:background} presented the RPL protocol, the Routing Table Falsification attack. \textcolor{black}{Section\ref{Sec:relatedwork} presented the literature of some RPL specific attacks.} Section \ref{Sec:systemadversary} presents the preliminary requirements, which are essential to design our proposed solution. Moreover, our proposed approach is presented in Section \ref{Sec:Proposed Solution}. Section \ref{Sec:exp} elaborates on the simulation parameters and presents an analysis of the network's performance based on various metrics. \textcolor{black}{Further, section \ref{Rev3} presents an encrypted variant of the proposed solution.} The final Section \ref{Sec:Conclusion} summarizes the conclusions drawn in the paper.

\section{Background} \label{Sec:background}
\subsection{Overview of the RPL protocol}

Routing Protocol for Low Power and Lossy Networks (RPL) operates on the principle of distance-vector and source routing. RPL aims to establish routes between sensor nodes within the topology. RPL creates a Directed Acyclic Graph (DAG) structure to manage the routing of data packets. The DODAG comprises sensor nodes that represent the devices in the network and the edges that represent the links between the devices. The DAG structure enables RPL to efficiently route packets even in networks with low bandwidth and high loss rates.
 % RPL RFC 6550 specifies RPL as a ``Proposed Standard" \cite{rfc6550}. It is specially designed for LLN because creating and maintaining network topology requires less energy. That is why RPL is considered an energy-efficient protocol\cite{gaddour2012rpl, SHARMA202312}.
The sensor nodes are structured to converge towards a single destination, known as Destination-Oriented Directed Acyclic Graphs (DODAG). The DODAG is an acyclic structure in which each sensor node forwards its data toward the sink. There may be multiple DODAGs running simultaneously in the network to achieve fault tolerance. The sink node of the DODAG is called the 6LoWPAN Border Router (6LBR). Figure \ref{fig:RPL} represents the overview of RPL. There are four control messages available in RPL for topology construction and maintenance: (a) DODAG Information Solicitation (DIS), (b) DODAG Information Object (DIO), (c) Destination Advertisement Object (DAO), and (d) DAO-Acknowledgment (DAO-ACK).

The new node broadcasts the DIS message to request DIO messages when it wants to join an existing network. Upon receiving a DIS message, a sink node or intermediate node responds with a DIO message.
The DIO message is periodically broadcast to advertise its existence and provide routing information to other nodes in the network. The DIO contains information about the DODAG Version, DODAG ID, rank, supported Objective Function (OF), and other parameters. The rank value serves to indicate the position of a node relative to the sink node. Upon receiving the rank in the DIO packet, the new node calculates its rank and manages the parent list accordingly \cite{goel2023cra}. The sensor node with the lowest rank value among the surrounding nodes is selected as the preferred parent node. The supported OF in the DIO message is used to calculate the rank. OF is used to select the best path based on the rank value and routing metrics.
The RPL has the availability of two modes, i.e., storing and non-storing modes. In storing mode, after selecting the preferred parent, the DAO message is unicast to the selected preferred parent for route registry in their routing tables, and the message is forwarded up to the sink. In the non-storing mode of RPL, the child unicasts the DAO message to the sink. The DAO-ACK message is used to acknowledge the DAO message. In RPL, a timer mechanism is present to dynamically adjust the transmission of control packets\cite{levis2011trickle} known as Trickle Timer.

\subsection{\textcolor{black}{RPL-specific attacks}}
\textcolor{black}{Mayzaud \textit{et al.} \cite{mayzaud2016taxonomy} proposed a taxonomy of routing attacks against the RPL protocol. Some of the RPL-specific attacks are discussed here:
\begin{itemize}
    \item Forwarding Misbehavior attack: The malicious node accepts data packets from neighboring or upstream nodes but deliberately fails to forward them to their destination, instead discarding them.
    \item DAO Inconsistency attack: A malicious node deliberately discards received data packets and responds with a forwarding error message, misleading the parent node into removing valid downward routes from its routing table.
\item Hatchetman attack: A malicious node modifies the source header of control packets and generates a large volume of invalid packets containing incorrect routing information, aiming to disrupt legitimate nodes.
\item DIO-Suppression attack: Malicious node deliberately block or suppress DIO messages, preventing legitimate nodes from receiving critical routing updates and causing disruptions in the network routing process.
\item Energy Depletion attack: A malicious node floods legitimate nodes with a large volume of packets, aiming to exhaust their energy resources.
\item Spam DIS attack: A malicious node generates and broadcasts a large number of DIS messages using spoofed or fake source identities, overwhelming the network and triggering unnecessary routing updates.
\item Sybil attack: A malicious node broadcasts DIS messages using numerous fake identities throughout the network in an attempt to manipulate and take control of the routing process.
\item Advanced Vampire attack: A malicious node alters the source routing information, causing legitimate nodes to reject the packet and respond with error messages, ultimately leading to routing instability and disruption of network services.
\end{itemize}}

\subsection{\textcolor{black}{Overview of PUF}}
\textcolor{black}{Physical Unclonable Function (PUF) is a hardware security mechanism that leverages the inherent manufacturing variations of physical devices to generate unique, unpredictable, and tamper-resistant identifiers. These identifiers are used for applications such as secure key storage, device authentication, and cryptographic operations \cite{al2023physical}. PUFs are essential for addressing security challenges inherent in IoT Networks. IoT devices are often resource-constrained and deployed in untrusted environments. PUFs provide a lightweight, hardware-based solution that enhances security without requiring significant computational overhead \cite{pu2022lightweight}.
There are several applications of PUFs in IoT: 
\begin{itemize}
    \item[1]  {Device Authentication}: A PUF generates a unique response for a given challenge based on the device's physical properties. This challenge-response mechanism ensures that only authorized devices can participate in the IoT network.
    \item[2] {Cryptographic Key Generation}: PUFs generate cryptographic keys dynamically from hardware properties, eliminating the need for storing sensitive keys in non-volatile memory.
    \item[3] {Secure Communication}: PUF-derived keys are used for encrypting data transmitted between IoT devices, ensuring confidentiality and integrity in communication.
    \item[4] {Anti-Tampering Measures}: The physical attack on a PUF typically alters the underlying hardware properties, invalidating the responses and alerting the system.
\end{itemize}}

\begin{figure}
    \centering
    \includegraphics[width=.5\textwidth]{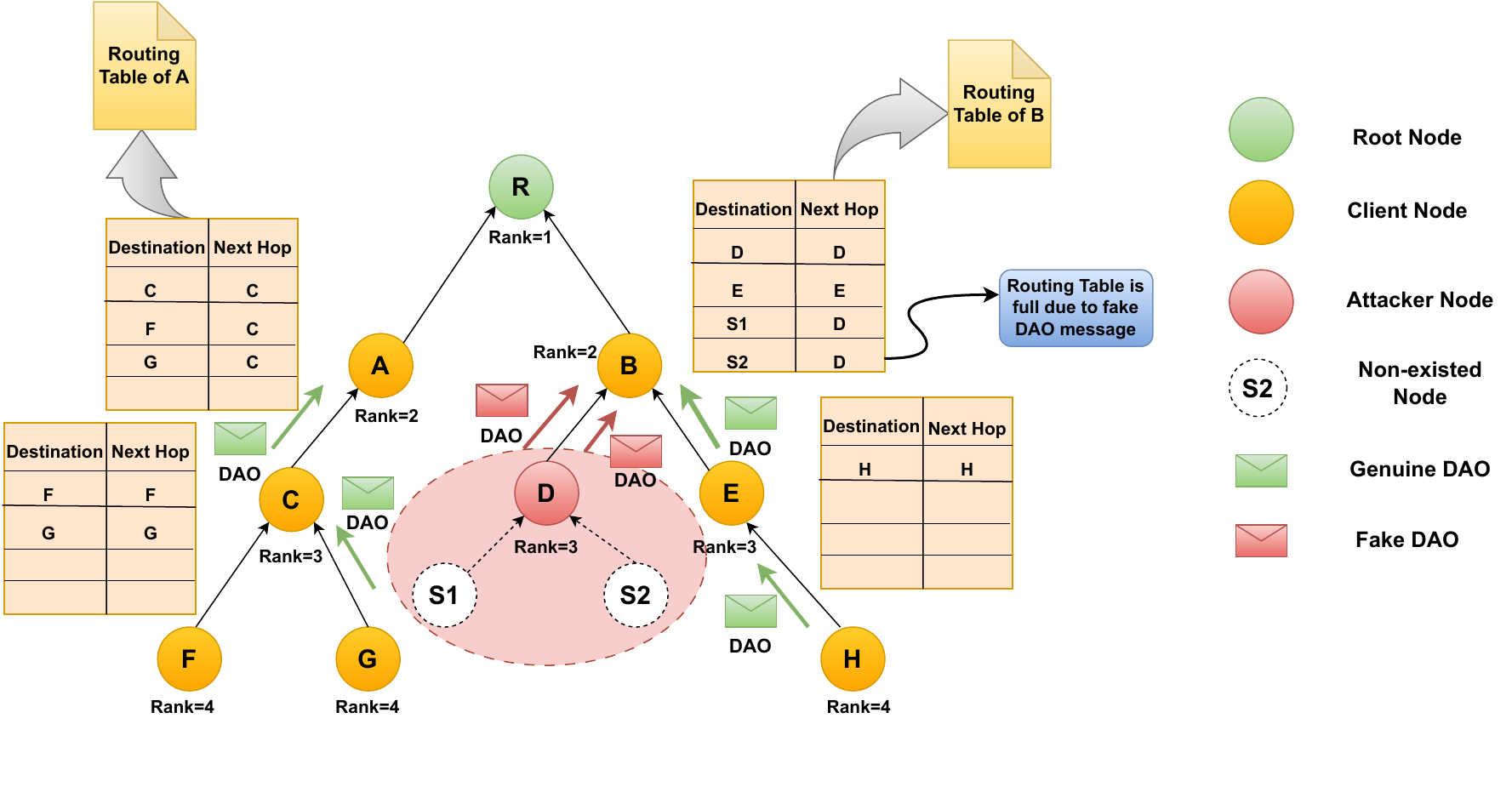}
    \caption{An Illustration of Routing Table Falsification Attack}
    \label{fig:RTP}
\end{figure}
\subsection{Routing Table Falsification Attack}

The DAO control message is utilized in RPL to establish downward routes from the sink node to the leaf nodes. RPL offers two modes: storing mode and non-storing mode. In non-storing mode, the child node unicasts the DAO message to the sink via the parent node, which is selected with the aid of OF. The parent node adds its address to the header and forwards the DAO message to the sink. Upon receiving the DAO, the sink node sends a DAO-ACK to the child node, which originated the DAO message. However, in the storing mode of RPL, the child node unicasts DAO messages directly to the parent. Upon the establishment of a new route between the parent and child nodes, the parent's routing table is updated with an entry for this route. Following this, the route must be registered at the sink node by forwarding the DAO message through the preferred parent node. In response to the DAO message, the sink node subsequently unicasts a DAO-ACK, which is forwarded to the child node through the downward routing path. The acknowledgment in the form of DAO-ACK confirms the registration of the child node at the sink. In this mode, the intermediate routers in the network maintain routing tables to store information about routes. The routing table of the standard RPL is exploited to mount the RTF attack. The intruder may compromise an insider node to perform this attack. By introducing an RTF attack in RPL, the malicious nodes forge the DAO message of the standard RPL to promote fake routes to the parent to disrupt the resource-constrained nature of sensor nodes.

The RTF attack is introduced in Mayzaud \textit{et al.} \cite{mayzaud2016taxonomy}. During an RTF attack, the malicious node is an internal node and serves as the child of the parent node. The malicious node unicast forged DAO messages to the parent to fill the routing table of the parent node. The impact of this attack on RPL networks has not been studied or documented yet \cite{DBLP:journals/csur/BangRKC23}.
Figure \ref{fig:RTP} represents the RTF attack in RPL. In the typical scenario of RPL, Nodes F, and G unicast DAO messages to the parent node C, and the entry is shown in the routing table of C. Node C forwards the DAO message of their child node to its parent node A for route registry in the routing table of A. This is the storing mode operation of the standard RPL. In the attack scenario, an insider node D is compromised to perform an RTF attack. The malicious node D unicasts the forged DAO message of Nodes S1 and S2 (non-existing nodes) to the parent node B. Node B registers the route entries of Nodes S1 and S2 by considering them as genuine nodes. Afterward, when Node E unicasts the DAO message of their child Node H, the routing table of B is filled by the non-existing routes unicast by the malicious node D. As a result, the routing table entries are full, thus blocking
the creation of routes towards new legitimate nodes (such as H) \cite{kamble2017security}. The attack impacts longer paths, increased network delays, higher packet drop rates, and even network congestion.
 \begin{table*}[]
\caption{\textcolor{black}{A comparison of various attacks with their countermeasure}}
 {
 \scalebox{0.9}
{
\begin{tabular}[c]{p{1.5cm} p{3cm} p{4cm} p{3.5cm} p{2cm} p{2.5cm}}\hline
 \small
	% 	\footnotesize
%	\label{Symbols}\\

		\textbf{Reference}&\textbf{Targeted attack} & \textbf{Description of attack}& \textbf{Defense approach} &\textbf{Mobility support}& \textbf{Type of validation}\\ \hline

%\endhead
Perazzo \textit{et al.} \cite{perazzo2017dio} & DIO Suppression attack & Malicious nodes selectively block or suppress DIO messages that cause legitimate nodes may not receive important routing updates, leading to routing disruptions. & No solution exist & No & Simulation. \\ \hline
Pu \textit{et al.} \cite{pu2018mitigating} & Forwarding misbehavior attack & A malicious node silently drop incoming data packets instead of forwarding them. & Monitoring based approach known as CMD & No & Simulation. \\ \hline

Pu \textit{et al.} \cite{pu2018mitigating}&   DAO Inconsistency attack &  A malicious
node intentionally drops the received data packet and replies the forwarding error packet to cause the parent node to discard valid downward routes from the routing table. & Threshold based approach & No & Simulation. \\ \hline

 Pu \textit{et al.} \cite{pu2018hatchetman}& Hatchetman attack & A malicious node alter the source header of the control packet and then produces a huge number of invalid packet with erroneous routes targeting legitimate nodes. & No solution exist & No & Simulation. \\ \hline
   Pu \textit{et al.} \cite{pu2019spam} & Spam DIS attack & A malicious node sends a flood of DIS messages with fake source identities. & No solution exist & No & Simulation.\\ \hline

Pu \textit{et al.} \cite{pu2019energy} & Energy Depletion attack & A malicious node sends excessive packets to legitimate nodes in order to drain the energy of the legitimate nodes. & Misbehaviour Aware Detction (MAD)  & No & Simulation.\\ \hline
 % Verma \textit{et al.} \cite{verma2020cosec}  & CoSec-RPL &  Copycat attack & Outlier Detection & Fast detection of the attacker and low false alarm rate & The security mechanism increases memory overhead on the network to maintain neighbor information. & No  \\ \hline
  Pu \textit{et al.} \cite{pu2020theil} & Advance Vampire attack & A malicious node manipulates source route, the legitimate nodes immediately drops the packet and reply with error message that cause routing instability and service disruption. & Theil-index based approach & No & Simulation. \\ \hline

    Pu \textit{et al.} \cite{pu2022lightweight}&Sybil attack & Malicious node multicast DIS messages with multiple fake identities within the network to gain control over the routing process.  & Bloom filter based approach known as liteSAD & No & Simulation as well as on testbed.\\ \hline
  
 % Mayzaud \textit{et al.} \cite{mayzaud2016taxonomy} &  Routing Table Falsification attack  & No defense mechanism & The paper introduces the name of attack & The author proposed the taxonomy of routing attacks. & The Routing Table Falsification attack is not studied or documented.\\ \hline
Our paper & Routing Table Falsification attack & A malicious node unicast forged DAO messages to the parent to overflow the routing table of the parent node.& PUF based security solution known as LiSec-RTF & Yes & Simulation as well as on testbed.  \\ \hline
\end{tabular}
}
 \label{rel}
}
\end{table*}

  \section{\textcolor{black}{Related Work}} \label{Sec:relatedwork}
 \textcolor{black}{Recognizing the significance of security concerns in RPL, various defense mechanisms have been proposed to address some of the RPL-based attacks.
Pu et al. proposed a solution against Forwarding misbehavior attack known as CMD. In this mechanism, each node monitors its preferred parent’s forwarding behavior, specifically tracking packet loss. The node compares its own observations with packet loss rates reported by one-hop neighbors to detect potential misbehavior. 
 Pu et al. addresses the DAO Inconsistency attack. The attacker node deliberately drops some data packets and replies with an error message to discard the entry of the node from the routing table. The proposed defense mechanism suggests that each parent dynamically adjusts the threshold of the error packet at a particular period.
The author proposed a security solution against Energy Depletion attack known as misbehavior-aware threshold detection. In this technique, each node monitors the number of packets received from its child nodes over a set time window and compares this count against a dynamically determined threshold to identify possible malicious behavior. 
The authors proposed a security system against Advance vampire attack. In which, each node track the destination MAC addresses of the incoming data packets and evaluates the evenness of the distribution using the Theil index, a statistical measure of inequality. Under normal conditions, destination addresses show a certain pattern, but during an advanced vampire attack, a malicious node randomly injects packets having fictitious MAC addresses. This causes a significant increase in distribution randomness, leading to an abnormally high Theil index value. The countermeasure activates to mitigate the attack by identifying and eliminating the malicious activity.
Pu et al. proposed a lightweight security solution against the sybil attack called LiteSAD, based on Bloom Filter. Bloom filters are probabilistic data structures that efficiently test whether an element is a member of a set. In this solution, each node maintains a record of legitimate neighbors using Bloom Filter. When a node receives multiple DIS messages and identities that do not match the filter, it suspects a Sybil attack. The authors validated LiteSAD through simulation as well as on real-world testbeds.}

\textcolor{black}{Although existing solutions can detect and mitigate a range of RPL-based attacks, they are not effective against RTF attack due to fundamental differences in attack characteristics. RTF attack exhibit distinct behavioral patterns compared to attacks such as Forwarding Misbehavior, DAO Inconsistency, Hatchetman, DIO Suppression, Energy Depletion, Spam DIS, and Advanced Vampire attacks. These existing attacks primarily target packet forwarding, control message manipulation, or energy exhaustion, whereas RTF attack specifically compromise the integrity and consistency of routing information within the RPL topology. As a result, current IDS mechanisms fail to detect RTF attack. Since a Sybil attack is a form of identity spoofing, existing countermeasures focus primarily on detecting anomalies in DIS messages exchanged during the network discovery phase. However, these solutions are limited in scope and are not designed to detect RTF attack, which involve manipulation of routing information during the parent registration phase. As a result, existing Sybil attack detection mechanisms are insufficient for identifying or mitigating RTF attack.}

\textcolor{black}{In this paper, we have focused only on the study of Routing Table Falsification attack, i.e., an identity spoofing attack. Moreover, we propose a defense mechanism to address this attack. Table \ref{rel}  provides a detailed comparison of the some RPL specific attacks in the literature.  LiSec-RTF aims to detect and prevent such falsified updates by integrating lightweight trust-based validation mechanisms suited for resource-constrained IoT environments.
The proposed LiSec-RTF mechanism is effective in detecting forged DAO control messages that mislead routing decisions. However, to address the above mentioned attacks, the proposed approach can be further enhanced and extended. Since sybil attack is a kind of identity spoofing attack, it can also be detected by LiSec-RTF. }

% \textcolor{black}{Moreover, we propose a defense mechanism to address this attack. Table \ref{rel}  provides a detailed comparison of the some RPL specific attacks in the literature.  LiSec-RTF aims to detect and prevent such falsified updates by integrating lightweight trust-based validation mechanisms suited for resource-constrained IoT environments.
% The proposed LiSec-RTF mechanism is effective in detecting forged DAO control messages that mislead routing decisions. However, to address the above mentioned attacks, the proposed approach can be further enhanced and extended.
%  }
\section{Preliminary}\label{Sec:systemadversary}
\begin{table}[!h]
	\centering
	% 	\footnotesize
	\caption{Notations and Definitions}
	\begin{tabular}{p{5cm}p{3cm}}
		\hline
		\textbf{Notation} & \textbf{Definition}\\ \hline
        \textit{$\mathcal Reserved$} & Reserved field of DAO message.\\   
		\textit{$P$} & Preferred parent node.\\
  \textit{$\mathcal{C}$ }& Child node.\\
		\textit{$Max_{node}$} & Maximum count of sensor nodes. \\ 
       
         \textit{$\mathcal{RT}$} & Routing Table.\\
         \textit{$\mathcal R_{i}$} & Response of sensor node i. \\
          \textit{$\mathcal L_{i}$} & License of sensor node i.\\
           \textit{$\mathcal{CH}$} & Challenge. \\ 
          \textit{$r_{i}$} & Calculate response of sensor node i. \\ 
         \textit{$DAO_{modified}$} & Modified DAO control packet. \\ 
		\textit{$\mathcal N_{bl}$} & Count of Blacklist nodes. \\ 
	  \textit{$\mathcal{B}$ $ \gets $ $ \textit{[}1,\dots,Max_{node} \textit{]}  $ } & Number of entries in the Blacklist table. \\ 
		\textit{$\mathcal{N}$ $ \gets $ $ \textit{[}1,\dots,Max_{node}\textit{]} $} & Number of entries in the Neighbor table \\ 
	\begin{tabular}[c]{@{}c@{}}\\ \textit{$\mathcal {PUF}_{data}$ $ \gets $ \textit{[}$ <\mathcal{S}_{i}>$, $ <\mathcal{CH}>$, $ <\mathcal R>$\textit{]}} \\{$i  = $ $ 1 $,\ldots, $Max_{node}$}\end{tabular} & Structure of PUF data at 6LBR. \\ 
 $\mathcal{RT}$ $ \gets $ $ \textit{[}R_{1}, R_{2},\dots,R_{n}\textit{]} $ & Structure of Routing Table. \\
		\begin{tabular}[c]{@{}c@{}}\\ \textit{$\mathcal{M}_i$ $ \gets $ \textit{[}$ < $$ Src_{ip}$>$\textit{]}$} \\{$i  = $ $ 1 $,\ldots, $Max_{node}$}\end{tabular} & Framework of malicious node in the Blacklist table. \\ 
		   $\oplus$ & exclusive OR \\
		\textit{$ Src_{ip}$}  & Source address of DAO. \\ \hline
		
	\end{tabular} 
 \label{tab:symbols}
\end{table}

\begin{figure*}
    \centering
    \includegraphics[width=.85\textwidth]{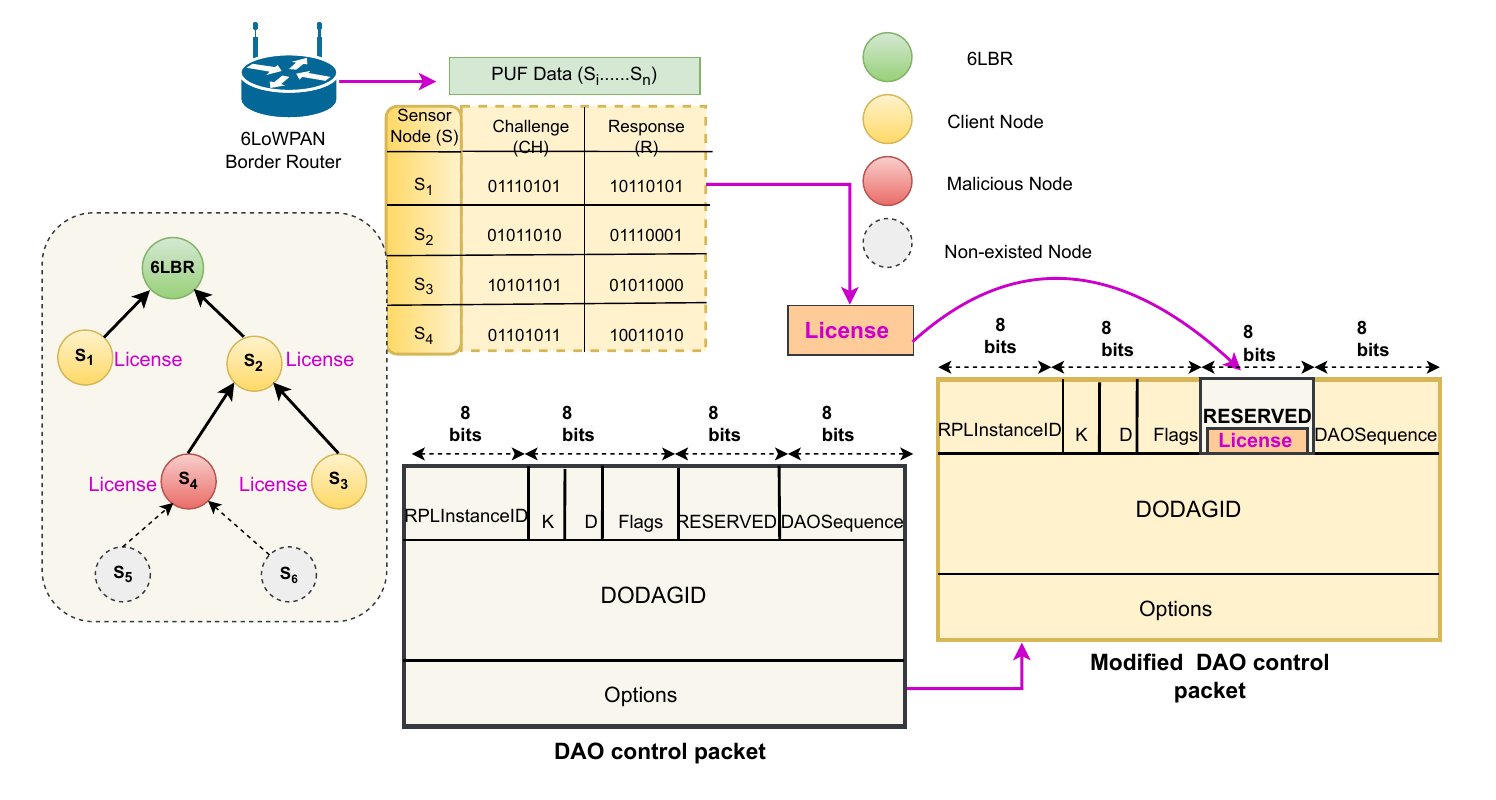}
    \caption{\textcolor{black}{System and Threat model}}
    \label{fig:s4}
\end{figure*}
\subsection{System and Threat Model} 

The system and threat model considered in this paper is depicted in Figure \ref{fig:s4}. Table \ref{tab:symbols} represents the notations and their corresponding definition for the proposed model. The assumption of the system model are outlined as follows:

% \subsection{System model}

\begin{itemize}

\item The study focuses on a PUF-enabled IoT network comprising a set $S$ = {$S_{1}$, $S_{2}$, $S_{3}$,..., $S_{n}$} of $n$ nodes. These nodes are IoT sensors characterized by limitations in communication range, memory capacity, processing capability, and battery capacity.

\item The sink node, also known as the 6LoWPAN Border Router (6LBR), possesses abundant resources. The 6LBR stores the PUF data ($\mathcal {CH-R}$) of each sensor node generated at the registration phase.
% \item For experimentation, the homogeneous nodes are considered, which send the DAO control packet to the root node for route registration and DAG maintenance.
\item \textcolor{black}{As shown in Fig. \ref{fig:s4},} the DAO message is modified ($DAO_{modified}$) to use the $8$-bit $\mathcal Reserved$ field to incorporate our proposed approach.
\item Each sensor node has its own License ($\mathcal L_{i}$) that is provided by the administrator using the node's PUF data ($\mathcal {CH-R}$).
% \item During the route registration phase, each sensor node registered at 6LBR.

\item The assumed system model is applicable in those applications where an administrator provides an authentication code before deployment, e.g., military weapons, Advanced Metering Infrastructure (AMI) \cite{cam2017applicability}.

% In this way, a sensor node is configured with its License key ($L_{i}$) and is unaware of the license keys of other sensor nodes.
This ensures the secure operation of LiSec-RTF.
\end{itemize}

% \subsection{Threat Model}
In the assumed network scenario, the malicious user is considered to have the following features.

\begin{itemize}
\item The malicious user can compromise the legitimate node and reprogram it to behave maliciously. \cite{bang2022embof,kaliyar2020lidl,challa2017secure}.
\item The malicious node is the insider node of the network. For experimentation purposes, the topology has a number of malicious nodes.
\item In the assumed IoT scenario, the malicious node pretends that it is the parent of several child nodes. The malicious node unicast the DAO packet with the falsified route to 6LBR through its parent.
\item The goal of a malicious user is to overload the routing table of the parent node with fake routes to create congestion on the network.
\item As shown in Figure \ref{fig:s4}, the insider node $S_{4}$ is registered in the network and has its License generated at the time of registration. Later on, it is captured by an attacker to mount an RTF attack.
\end{itemize}

\textcolor{black}{This work focuses on  an non-encrypted lightweight authentication for applications like smart agriculture, IIoT (Machine Monitoring), smart street lightening, wildlife tracking, smart parking systems, and conservation, where the data itself may not be highly sensitive (i.e. confidentiality is not a major concern) and has a need of low computational overhead based solution, requires only simple device identity verification due to regulatory compliance, or have open-field deployment. Such applications demand low-overhead, low-cost, simple deployment, and limited security. However, for applications which require more security against smart adversary and eavesdropping attack a encrypted variant of the proposed solution is discussed further in Section \ref{Rev3}.}

\section{Proposed Approach} \label{Sec:Proposed Solution}
\begin{figure}
    \centering
    \includegraphics[width=.5\textwidth]{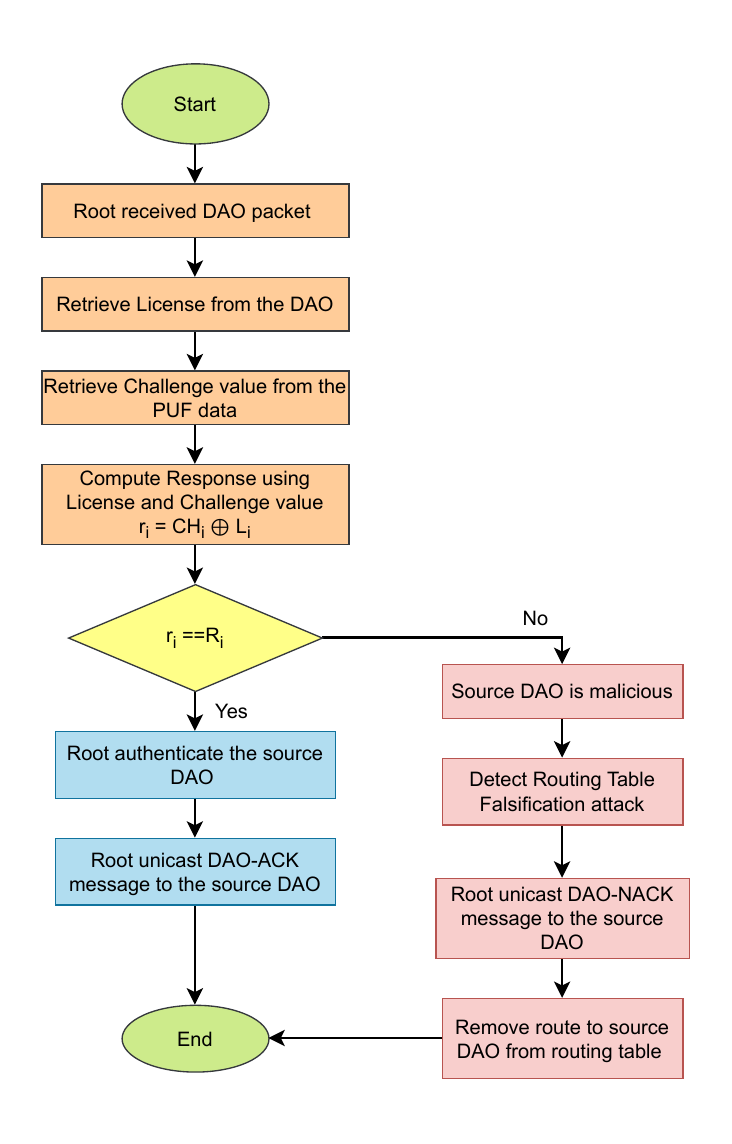}
    \caption{Overview of proposed approach}
    \label{fig:flow}
\end{figure}

\begin{figure*}
    \centering
    \includegraphics[width=.85\textwidth]{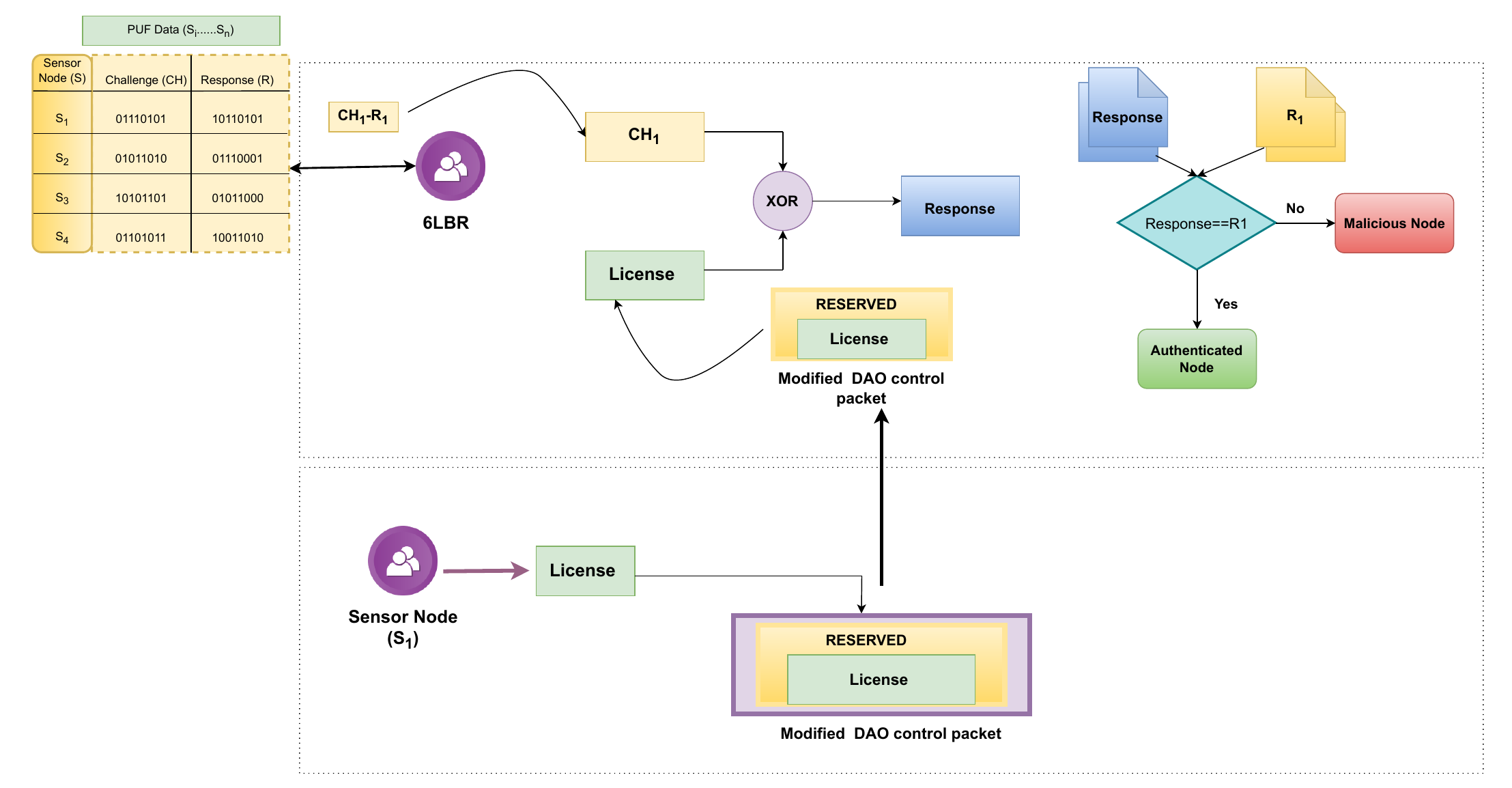}
    \caption{\textcolor{black}{Proposed Architecture of LiSec-RTF}}
    \label{fig:LISEC}
\end{figure*}

The current specification of RPL lacks a mechanism to counteract the RTF attack. We have proposed an approach named LiSec-RTF to address this limitation of RPL. \textcolor{black}{Figure \ref{fig:flow} illustrates the overview of our proposed approach.} In the proposed solution, we employ a $DAO_{modified}$ message, for authentication of nodes at 6LBR. The $\mathcal Reserved$ field encapsulates the Licence in the modified DAO ($DAO_{modified}$) packet. The License specifies a sequence of bits achieved by eXclusive OR (XOR) of the PUF data of a sensor node. \textcolor{black}{The proposed LiSec-RTF leveraging strong PUFs to generate a large number of challenge-response pairs which makes them suitable for applications like authentication and anti-cloning mechanisms in large-scale networks.} When a child node unicasts the $DAO_{modified}$ to the 6LBR, the 6LBR extracts the License from the $\mathcal Reserved$ field of the $DAO_{modified}$ for validation. Afterward, the 6LBR validates the License, whether it is a genuine License generated at the time of registration using their PUF data or a fake License provided by the malicious node to overload the routing table through fake routes. \textcolor{black}{Figure \ref{fig:LISEC} illustrates the proposed architecture of LiSec-RTF.} The License generation procedure is described in Section \ref{SEC:LI}.

The algorithm LiSec-RTF is illustrated in Algorithm \ref{Algo 1}. If the License is validated, it confirms that the node unicasted the DAO packet with a genuine route. For the generation of the License at the time of registration, we used PUF information of a node \cite{pu2022lightweight}.
 \begin{figure}
    \centering
    \includegraphics[width=.5\textwidth]{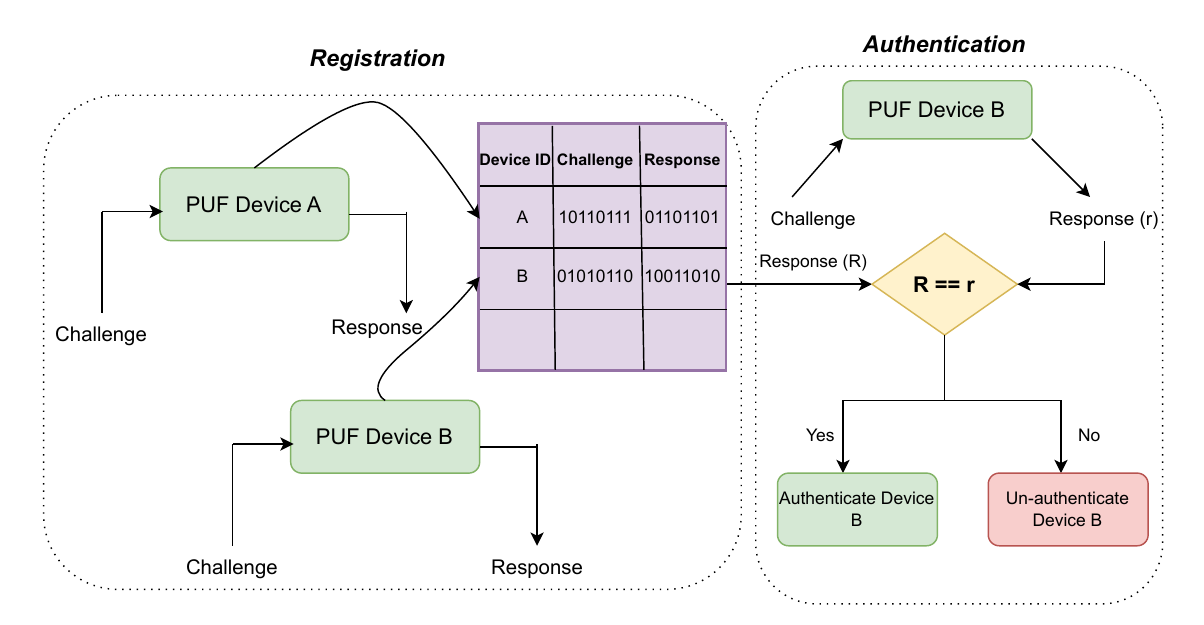}
    \caption{Overview of PUF for device authentication}
    \label{fig:puf}
\end{figure}
\subsection{Physical Unclonable Function (PUF)}
A Physical Unclonable Function (PUF) is a hardware security primitive that uses the physical variations available in Integrated Circuits (ICs) to create a unique authentication code \cite{Guajardo2011}. These physical variations are nearly impossible to replicate precisely to generate an identifier that resembles a fingerprint for every distinct chip. When the binary sequence of inputs named ``Challenge" is applied to the PUF, it will produce the corresponding output named ``Response". The Challenge-Response pair of each IC is unique due to the feature of PUF, and no two IC will produce an identical response for the same Challenge.
 
Device authentication based on PUF can be implemented in two phases:
(1) Registration; (2) Authentication. In the registration phase, Devices A and B are registered in the trusted environment, as depicted in Figure \ref{fig:puf}. In the next phase to authenticate Device B, the same Challenge is passed as the input to Device B, and it will produce the Response. If both responses are the same, we consider that Device B is authenticated. Otherwise, it is a malicious device by keeping the identity of Device B.

\begin{figure}
    \centering
    \includegraphics[width=.5\textwidth]{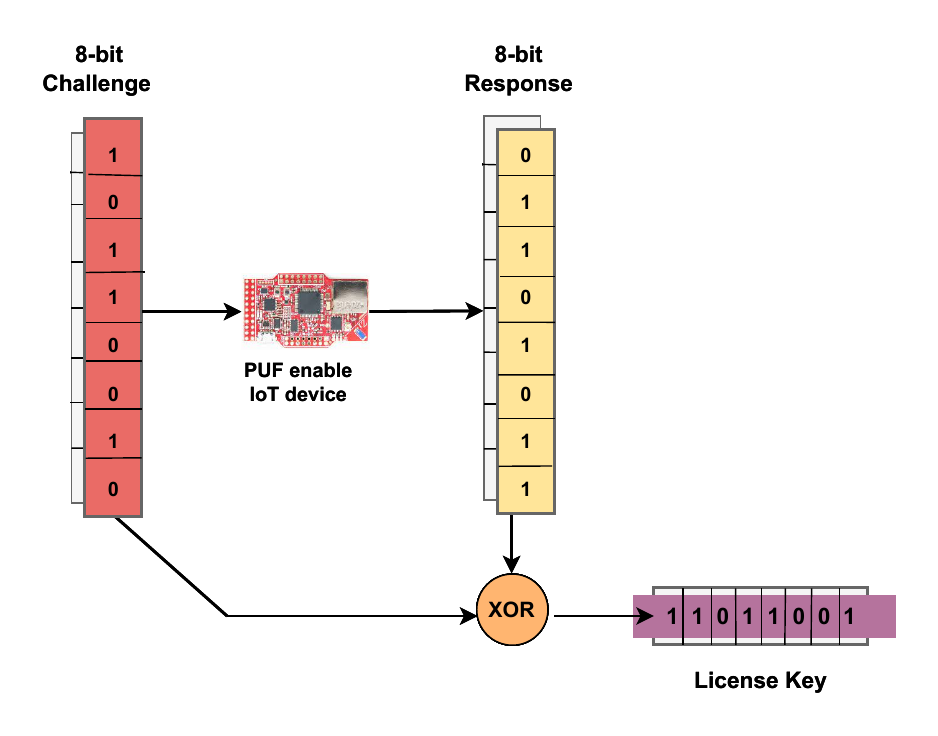}
    \caption{License Generation Procedure}
    \label{fig:lk}
\end{figure}

\subsection{License Generation Procedure} \label{SEC:LI}
Figure \ref{fig:lk} illustrates the License generation  procedure of a PUF-enabled sensor device. When a $8$-bit random value called as Challenge ($\mathcal {CH}$) is given as input to the PUF-enabled sensor node, the $\mathcal {CH}$ is passed to the function of PUF which produces an $8$-bit output known as Response ($\mathcal R$). This process is represented with the help of Eq. \ref{eq1}. 

\begin{align}  \label{eq1}
       \mathcal R = f_{PUF}(\mathcal {CH})
    \end{align} 

License ($ \mathcal L_{i}$) is generated at the time of registration and represented by Eq. \ref{eq2}. Each sensor node is configured with the generated License during node deployment phase of RPL.

\begin{align}  \label{eq2}
     \mathcal L_{i} = \mathcal{CH}_{i} \oplus \mathcal R_{i}
    \end{align} 

% \textcolor{black}{ The $ \mathcal L_{i}$ is encrypted using a lightweight encryption algorithm such as Advanced Encryption Standard (AES) or Elliptic Curve Cryptography (ECC) before transmission.  
% The encrypted License can be represented as:  
% \begin{align}  \label{eqs}
%     \mathcal{L}_{i}^{enc} = Encrypt(K_{shared}, \mathcal{L}_{i})  
% \end{align}  
% where \( K_{shared} \) is a pre-shared symmetric key or dynamically generated session key between the sensor node and the 6LBR.  }

\subsection{Detection mechanism of LiSec-RTF}
Fig. \ref{fig:LISEC} represents the framework of LiSec-RTF. When the 6LBR receives the modified version of DAO message ($DAO_{modified}$). It extracts the License ($\mathcal L_{i}$) from the $\mathcal Reserved$ field of the DAO message.
% \textcolor{black}{Upon receiving the modified DAO message, the 6LBR decrypts the \(\mathcal{L}_{i}^{enc}\) before proceeding with validation. 
For the purpose of authentication, 6LBR extracts the Challenge ($\mathcal {CH}_{i}$) from the PUF data ($\mathcal{PUF}_{data}$) of that sensor node and decrypts the \(\mathcal{L}_{i}^{enc}\) from the $DAO_{modified}$. 
% \begin{align}  \label{eq3}
%     r_{i} = \mathcal{CH}_{i} \oplus Decrypt(K_{shared}, \mathcal{L}_{i}^{enc})  
% \end{align} } 
\begin{align}  \label{eq3}
      r_{i} = \mathcal{CH}_{i} \oplus \mathcal L_{i}
    \end{align} 

In Eq.\ref{eq3}, the calculated response ($ r_{i}$) is the XOR of $\mathcal {CH}_{i}$ and $\mathcal L_{i}$. At the 6LBR, if the calculated response ($ r_{i}$) is equal to the Response ($\mathcal R_{i}$) that is stored in their ($\mathcal{PUF}_{data}$), then we state that the License ($\mathcal L_{i}$) is validated and the source address ($ Src_{ip}$) of the DAO sender is authenticated; otherwise, it is a fake License generated by the malicious node for registering the fake routes to overload the routing table of the parent node. LiSec-RTF is depicted in Algorithm \ref{Algo 1}.

\begin{algorithm}
 % \flushleft   
\small
% \scriptsize
\caption{Pseudocode of LiSec-RTF}
\label{Algo 1}
\begin{algorithmic}[1]

\Procedure{Initialization}{}
 % \State $pkt[type]$ \Comment{Type of control packet}
     \State $\mathcal {PUF}_{data}$ $ \gets $ $ \textit{[}1,\dots,Max_{node}\textit{]} $ 
    \State $\mathcal{B}$ $ \gets $ $ \textit{[}1,\dots,Max_{node}\textit{]} $ 
     \State $\mathcal{N}$ $ \gets $ $ \textit{[}1,\dots,Max_{node}\textit{]} $ 
     \State $\mathcal{RT}$ $ \gets $ $ \textit{[}R_{1}, R_{2},\dots,R_{n}\textit{]} $
       \State $\mathcal L_{i} = \mathcal{CH}_{i} \oplus \mathcal R_{i} $ \Comment{Eq. \ref{eq2}}
\EndProcedure

\Procedure{\textcolor{black}{On\_${Sender}$}}{$pkt$}
  \If{($pkt[type]$ == $pkt[DAO_{modified}$])}
  \State $\mathcal Reserved$ $\xleftarrow{}$~ $ \mathcal{L}_{i}$
  \State Send $DAO_{modified}$ to $P$
  \EndIf
\EndProcedure
\Procedure{\textcolor{black}{On\_${Receiver}$}}{$pkt$}
    \If{($pkt[type]$ == $pkt[DAO_{modified}$])}
    \State Receive $DAO_{modified}$ from $\mathcal{C}$
   \State Add route to $\mathcal{C}$ in $\mathcal{RT}$ \Comment{Add route in routing table}
    \State Forward $DAO_{modified}$ to 6LBR
    \ElsIf{($pkt[type]$ == $pkt[\mathcal DAO-ACK$])}
     \State Forward $\mathcal DAO-ACK$ to $\mathcal{C}$
       % \State Continue 
      \ElsIf{($pkt[type]$ == $pkt[\mathcal DAO-NACK$])}
        \State Forward $\mathcal DAO-NACK$ to $\mathcal{C}$
     %  \If {($\mathcal DAO-NACK$)}
     \State $\mathcal{B}$  $\xleftarrow{}$~ $ Src_{ip}$ 
       \State Remove route to $\mathcal{C}$ from 
    $\mathcal{RT}$
      \State Remove $ Src_{ip}$ from 
      $\mathcal{N}$
      \State $\mathcal N_{bl}++$
    \Else
    % \State Continue 
    %  \EndIf
    %\Else
    \State Do nothing
    \EndIf
\EndProcedure
\Procedure{\textcolor{black}{On\_6LBR}}{$pkt$}
 \If{($pkt[type]$ == $pkt[DAO_{modified}$])}
 \State $\mathcal L_{i}$  $\xleftarrow{}$~ $\mathcal Reserved$
 \State $\mathcal {(CH-R)}_{i}$  $\xleftarrow{}$~ $\mathcal {PUF}_{data}(1,\dots,Max_{node})$
 \State $ r_{i} = \mathcal{CH}_{i} \oplus \mathcal{L}_{i}  \cdots$ by Eq. \ref{eq3} 
  \If {($\mathcal{R}_{i}$ \textbf{equals} $r_{i}$)}
   \State unicast $\mathcal DAO-ACK$
    \Else
    \State unicast $\mathcal DAO-NACK$
    \EndIf
    \EndIf
   \EndProcedure
% \Procedure{At\_$P$}{$\mathcal DAO-ACK/NACK$\_Receive}
%      \State Forward $\mathcal DAO-ACK/NACK$ to $\mathcal{C}$
%        \If {($\mathcal DAO-NACK$)}
%      \State $\mathcal{B}$  $\xleftarrow{}$~ $ Src_{ip}$ 
%        \State Remove $\mathcal{C}$ from 
%     $\mathcal{RT}$
%       \State Remove $ Src_{ip}$ from 
%       $\mathcal{N}$
%       \State $\mathcal N_{bl}++$
%     \Else
%     \State Continue 
%      \EndIf
% \EndProcedure

\end{algorithmic}
\end{algorithm}

\subsection{\textcolor{black}{Description of LiSec-RTF}}
\textcolor{black}{The Algorithm \ref{Algo 1} illustrates the pseudocode for LiSec-RTF, which is integrated into the DAO processing function within the $rpl-icmp6.c$ file. The DAO control message plays a pivotal role in registering routes at the 6LBR node and maintaining the network topology. LiSec-RTF is triggered whenever the sender unicasts a DAO message to the root node via its preferred parent to facilitate route registration and topology maintenance.}
% The Initialization procedure in Algorithm \ref{Algo 1} performs the following tasks:
% \begin{itemize}
%     \item Initialization of PUF data at the 6LBR.
%     \item Initialization of Neighbor Table and Blacklist Table.
%     \item Initialization of Routing Table on each non-leaf node.
%     \item The 
% \end{itemize}

\textcolor{black}{The Algorithm \ref{Algo 1} contains three procedures. Procedure $1$ is On\_${Sender}$, which denotes the Sender procedure. In this procedure, the sender is the child node, which prepares the $DAO_{modified}$ packet by inserting the $\mathcal{L}_{i}$ in the $\mathcal Reserved$ field of the DAO and sending it to the preferred parent $P$. The second procedure is On\_${Receiver}$, which denotes the receiver procedure. In this procedure, if $pkt[type]$ is $DAO_{modified}$, then it receives the packet from the child node, adds the route to the child in the routing table of $P$, and forwards the packet to the 6LBR. Other than this, if $pkt[type]$ is DAO-ACK, it states that it receives the DAO-ACK from 6LBR and forwards it to the child node $\mathcal{C}$. When $pkt[type]$ is DAO-NACK, it states that it receives the DAO-NACK from 6LBR. The parent node forwards this packet to the child, inserts the source address in the blacklist table, and removes the route to the child from the routing table and neighbor table. In the last procedure, On\_6LBR, if $pkt[type]$ is $DAO_{modified}$, it receives the packet from $P$. The 6LBR extracts $\mathcal{L}_{i}$ from the $\mathcal Reserved$ field of the DAO. It extracts the child node ${PUF}_{data}$. The 6LBR calculates $ r_{i}$ using Eq. \ref{eq3}. If both response values are the same, then it unicasts the DAO-ACK; otherwise, it unicasts DAO-NACK to the child through $P$.
}
\subsection{\textcolor{black}{Mathematical formulation of LiSec-RTF}}
This section considers the assumed system and threat model as shown in Figure \ref{fig:s4}. The topology consists of a 6LBR and four sensor nodes ($S_{1}, S_{2}, S_{3}$ and $S_{4}$). The PUF data ($\mathcal{PUF}_{data}$) of these sensor nodes is deployed at 6LBR by using Eq. \ref{eq1}. The sensor node is configured with its License at the phase of deployment by using Eq. \ref{eq2}.

\textcolor{black}{Each sensor node $S_i$ has a unique PUF response, denoted as $\mathcal{PUF}_{data}$, represented as $(CH - R)_i$.}
% \begin{itemize} 
%    \item  \textcolor{black}{\textbf{License Generation:}
% Each node computes its License $L_i$ using the XOR operation by using Eq. \ref{eq2}.}
%  \textcolor{black}{\item \textbf{Transmission of $DAO_{modified}$:}
% The computed License $L_i$ is inserted into the $\mathcal Reserved$ field of the $DAO_{modified}$ message and unicasted to the 6LBR through the parent node.}
%  \textcolor{black}{\item \textbf{Verification at 6LBR:}
% Upon receiving the $DAO_{modified}$ message, the 6LBR retrieves $L_i$ and verifies it using the stored PUF data by using Eq. \ref{eq3}.}
% \textcolor{black}{\item \textbf{Authentication Condition:}
% The authentication of the sensor node $S_i$ is successful if:
% \begin{equation}
%     R_i = r_i
% \end{equation}}
% \end{itemize}

\begin{enumerate}
 \item \textcolor{black}{ The License $L_i$ of a sensor node is generated using $\mathcal{PUF}_{data}$:
  \begin{equation} \label{eqe}
       L_i = CH_i \oplus R_i
  \end{equation}}

  \item \textcolor{black}{ The sensor node unicasts $L_i$ in the $DAO_{modified}$ message.}

  \item \textcolor{black}{ Upon receiving the $DAO_{modified}$ message, the 6LBR extracts $L_i$ and uses its stored $CH_i$ to compute:
  \begin{center}
      $ r_i = CH_i \oplus L_i$
  \end{center}}
  
  \textcolor{black}{Substituting $L_i$ from Eq. \ref{eqe}:
  \begin{center}
      $ r_i = CH_i \oplus (CH_i \oplus R_i)$
  \end{center}}

  \textcolor{black}{Using the XOR properties:
  \begin{center}
       $A \oplus A = 0, \quad \text{and} \quad A \oplus 0 = A$
  \end{center}}

  \textcolor{black}{Applying this to simplify:
  \begin{center}
     $  r_i = (CH_i \oplus CH_i) \oplus R_i$
  \end{center}
  \begin{center}
      $ r_i = 0 \oplus R_i$
  \end{center}
  \begin{center}
       $r_i = R_i$
  \end{center}}
\end{enumerate}
\textcolor{black}{Since $r_i$ (computed at 6LBR) equals $R_i$ (original response stored at 6LBR), the authentication condition $R_i = r_i$ holds \textbf{true}.
The proof establishes that if the $DAO_{modified}$ message is not tampered with, the authentication check $R_i = r_i$ will always be \textbf{true} due to the properties of XOR. However, if a malicious node sends an incorrect $L_i$, the computed $r_i$ at 6LBR will not match $R_i$, leading to authentication failure.}

\textcolor{black}{The main feature of this mitigation approach is to authenticate each sensor node at the 6LBR. The purpose of using PUF is to identify malicious nodes, as PUF generates a device's unique bit pattern (Response). The 6LBR sends an arbitrary bit pattern (Challenge) to the PUF-enabled device, which then produces an output bit pattern (Response). If the Response matches the stored Response, it confirms that the particular node is authentic. Therefore, if an attacker unicasts a DAO message with fake identities, the 6LBR can detect the malicious node and respond with a NACK.
}

\textcolor{black}{In a brute-force attack scenario, an attacker may try all possible License combinations to gain authentication at the root node. Our proposed solution is secure against brute-force attacks, provided that a sufficiently large License size is used (16, 32, 64, 128 bits). In this paper, we use an 8-bit License stored in the unused RESERVED field of the DAO message to demonstrate how the proposed approach can be implemented in an RPL-based Contiki-NG environment. However, for better security against brute-force attacks, we have the option to store the License in the OPTIONS field of the DAO, which supports storing variable-sized information. Therefore, depending on the security requirements, a network administrator has the flexibility to utilize either the OPTIONS field or the RESERVED field to store Licenses for resisting brute-force attacks, based on the application or deployment scenario.}
\subsection{\textcolor{black}{Analysis of LiSec-RTF}}
\textcolor{black}{The $\mathcal {(CH-R)}_{1}$ for $S_{1}$ node is $(01110101 - 10110101)$. By using Eq. \ref{eq2}:
\begin{center}
         $\mathcal{L}_{1}$ = $(01110101~\oplus~ 10110101)$   \\
         $\mathcal{L}_{1}$ = $11000000$\\
\end{center}
The  $\mathcal{L}_{1}$ is inserted in the $\mathcal Reserved$ field of the $DAO_{modified}$ message and  unicast to the 6LBR through the parent node ($P$). When the $DAO_{modified}$ received at the 6LBR, it extracts the $\mathcal{L}_{1}$ from the $\mathcal Reserved$ field. The 6LBR also extracts the $\mathcal{PUF}_{data}$ $\mathcal {(CH-R)}_{1}$ of the $S_{1}$ node. By using Eq. \ref{eq3}:
\begin{center}
         $r_{1}$ = $(01110101~\oplus~ 11000000)$   \\
         $r_{1}$ = $10110101$\\
\end{center}}

\textcolor{black}{In this example, as we refer line number $36$ of Algorithm \ref{Algo 1} is true i.e., $(\mathcal R_{1} equals~ r_{1})$. Then, we state that sensor node $S_{1}$ is authenticated. Alternatively, the malicious node unicasts the DAO packet with the falsified route.}
\begin{table}[!h]
\centering
% 	\footnotesize
\caption{SIMULATION SCENARIO}
\begin{tabular}{p{3.5cm}p{4.2cm}}
\hline
\textbf{Parameters} & \textbf{Values}\\ \hline
Simulator & Cooja on Contiki-NG \\
Radio Model & Unit Disk Graph Medium (UDGM)\\ 
Mobility model & Random Waypoint Mobility Model\\
Size of Grid & $200m \times 200m$ \\ 
Objective function & Minimum Rank  with Hysteresis Objective Function (MRHOF)\\
Range of transmission & $50m$\\
Time of simulation & $1800s$ \textcolor{black}{(30 minutes)}   \\
Node type & z1 mote \\
Speed of node & 1-2 m/sec \\
Size of data packet & $30$~Bytes \\
Number of 6LBR node & $1$ \\
Number of client nodes & $29$ \\ 
Number of malicious nodes & $1$, $2$, $3$\\
\hline
\end{tabular}
\label{tab:sim-par}
\end{table}

\section{Performance Evaluation} \label{Sec:exp}
\subsection{Experimental Setup}
\textcolor{black}{In the simulations, the RTF attacker is implemented by making modifications to the existing file of rpl-icmp6.c in Contiki-NG. In the file, the attacker node unicast DAO message to its parent node with fake ip address in order to fill the routing table of its parent node.} LiSec-RTF is implemented in Contiki-NG operating system. We evaluated the performance of the LiSec-RTF in Cooja simulator. The proposed solution is implemented by making modifications to the existing files within the rpl-classic mode of the Contiki-NG. \textcolor{black}{The proposed solution can detect the identity spoofing attacks (A node impersonates another to mislead routing decision).} Table \ref{tab:sim-par} shows the details of simulation parameters. We have repeated the experiments with $10$ different random seeds for statistically accurate results. Subsequently, we utilized the mean values of these results along with their errors, computed at a $95$\% confidence interval, to mitigate any potential biases in the findings. 

\subsection{Evaluation metrics}
We employed Packet Delivery Ratio (PDR), Average End-to-End Delay (AE2ED), and Average Power Consumption (APC) to investigate the influence of an RTF attack on RPL and to verify the efficacy of the LiSec-RTF in both static and mobile environments.
Additionally, the memory overhead of the LiSec-RTF on the Zoletria z1 mote are also examined. These measures are described as follows:
\subsubsection{PDR}
        It represents the proportion of packets received successfully at 6LBR compared to the number of packets sent by each sensor node. It can be quantified using the Eq. \ref{eq7}
        
\begin{equation}\label{eq7}
     PDR =   \frac{D_{receieved}}{\sum_{i=1}^{N}D_{sent_{i}}}
     \end{equation}
     
where $D_{received}$ represents the total count of data packets received at the sink, and $D_{sent_{i}}$ denotes the total count of data packets sent by client node $i$.
    
\subsubsection{AE2ED}
AE2ED refers to the average time taken for data packets to travel from the client node to the 6LBR in a network. It is represented by Eq. \ref{eq8} 

\begin{equation}\label{eq8}
     AE2ED =    \frac{\sum_{i=1}^{N}D_{received_{i}}}{D_{N}}
     \end{equation}

where $D_{received_{i}}$ denotes the cumulative number of packets received by each client node $i$, while $D_{N}$ signifies the time delay experienced by the data packet.
  
\subsubsection{APC}
It refers to the average power consumed by each client in a specified time. Eqs. \ref{eq11} and \ref{eq12} denote energy and power, respectively. 

\begin{equation}\label{eq11}
   Energy(mJ) = (TX + RX + CPU + LPM)
   \end{equation}

TX represents the transmission, RX represents the receiving, LPM represents the low power mode, and CPU represents CPU time 
\cite{verma2020security}.

\begin{equation}\label{eq12}
       Power(mW) =\frac{Energy}{Tst}
\end{equation}
   
where $Tst$ represents the total simulation time in seconds.

\subsection{Results and Discussion} \label{Sec:res}
In this section, we analyze the performance of the network through simulation. \textcolor{black}{For comparison, we have chosen $RPL$, $RPL_{Under~Attack}$, and LiSec-RTF. Where RPL denotes the standard RPL, it serves as a baseline for performance comparison, representing the unaltered behavior of the RPL in ideal scenarios. $RPL_{Under~Attack}$ denotes the standard RPL that is being attacked by the insider node. The idea behind using this benchmark is to simulate RTF attacker nodes, which attempt to degrade network performance. LiSec-RTF refers to the secure version of RPL designed to counteract the RTF attack.} The performance of these three scenarios has been validated on PDR, AE2ED, and APC metrics. 

\subsubsection{Analysis on PDR}
PDR serves as a fundamental metric for evaluating the network's performance, reliability, and efficiency by successfully delivering data packets from client nodes to the sink node. We calculate PDR using Eq. \ref{eq7}. To measure the effect of an RTF attack in an IoT network, we have considered three cases: $RPL$, $RPL_{Under~Attack}$, and LiSec-RTF. Figures \ref{fig:spdr} and \ref{fig:mpdr} depict the values of PDR obtained while varying the count of malicious nodes in a static as well as mobile environment. In standard RPL ($RPL$), the sensor nodes are legitimate. The mean value of PDR in the case of $RPL$ is $1$ in static and $0.238$ in mobile scenarios. The value of PDR is lesser in mobile scenarios because sensor nodes may move unpredictably. The routes between mobile nodes may be less stable due to node mobility, leading to frequent route changes in network topology. In the case of an RTF attack ($RPL_{Under~Attack}$), the mean value of PDR is approximately $0.55$ to $0.71$ in the static scenario and $0.03$ to $0.05$ in a mobility environment while varying the number of malicious nodes.
By examining both scenarios, it becomes evident that the attack has a substantial impact on the mobile environment because the routes of the sensor nodes change very frequently in the network topology. When we simulate our proposed defense mechanism, the average value of PDR is $0.98$ in the static scenario and $0.19$ to $0.23$ in the mobile scenario.
As seen in Figures \ref{fig:spdr} and \ref{fig:mpdr}, our proposed approach, LiSec-RTF, increased the PDR as the standard RPL in static as well as in the mobile environment as it detects fake DAO packets unicast by the malicious insider nodes.

% \begin{figure*}[h]
% \centering
%   \begin{subfigure}[b]{.5\linewidth}
%     \centering
%     \includegraphics[width=0.7\textwidth]{pdrs.eps}
%     \caption{PDR in static scenario}
%     \label{fig:spdr}
% \end{subfigure}%   
%   \begin{subfigure}[b]{.5\linewidth}
%     \centering
%     \includegraphics[width=0.7\textwidth]{pdrm.eps}
%     \caption{PDR in mobile scenario}
%     \label{fig:mpdr}
%  \end{subfigure}%  
%  \caption{PDR obtained in different scenarios}
%  \end{figure*}

\subsubsection{Analysis on AE2ED}
\begin{figure}[h!]
\centering
 \begin{subfigure}[b]{.55\linewidth}
    \centering
    \includegraphics[width=\textwidth]{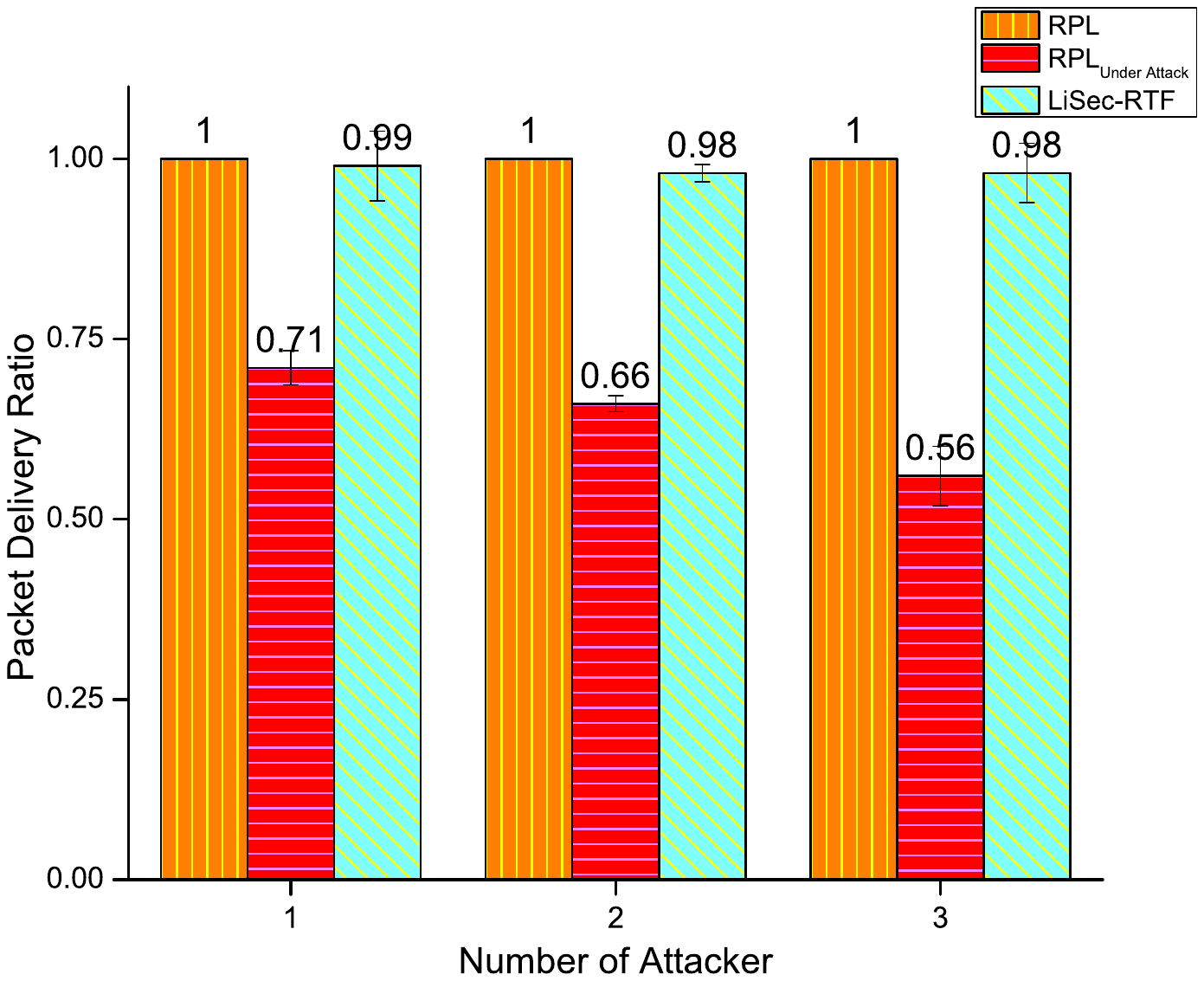}
    \caption{PDR in static scenario}
    \label{fig:spdr}
\end{subfigure}%   
  \begin{subfigure}[b]{.55\linewidth}
    \centering
    \includegraphics[width=\textwidth]{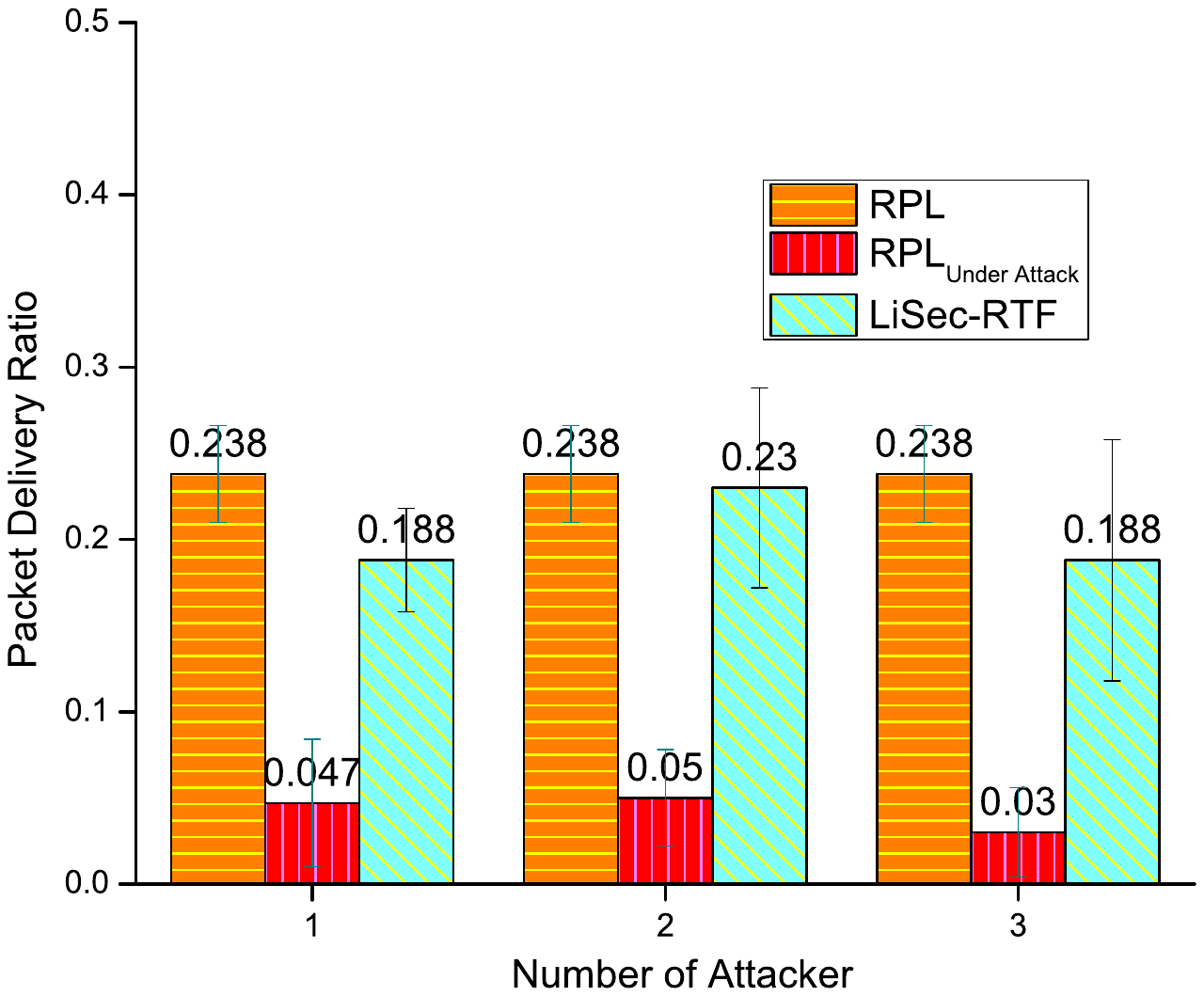}
    \caption{PDR in mobile scenario}
    \label{fig:mpdr}
 \end{subfigure}\\%  
  \begin{subfigure}[b]{.55\linewidth}
    \centering
    \includegraphics[width=\textwidth]{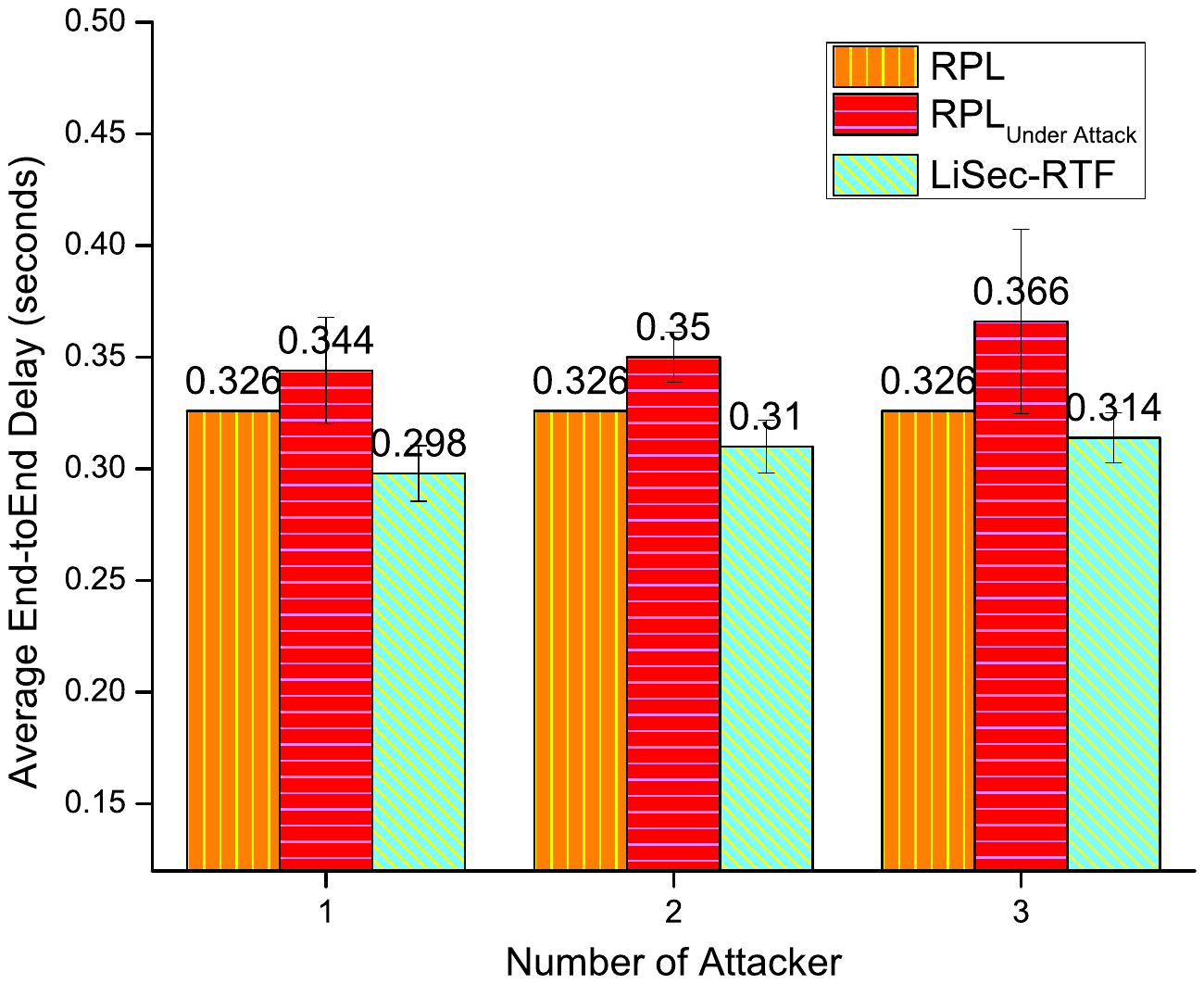}
    \caption{AE2ED in static scenario}
    \label{fig:se2e}
 \end{subfigure}%   
  \begin{subfigure}[b]{.55\linewidth}
    \centering
    \includegraphics[width=\textwidth]{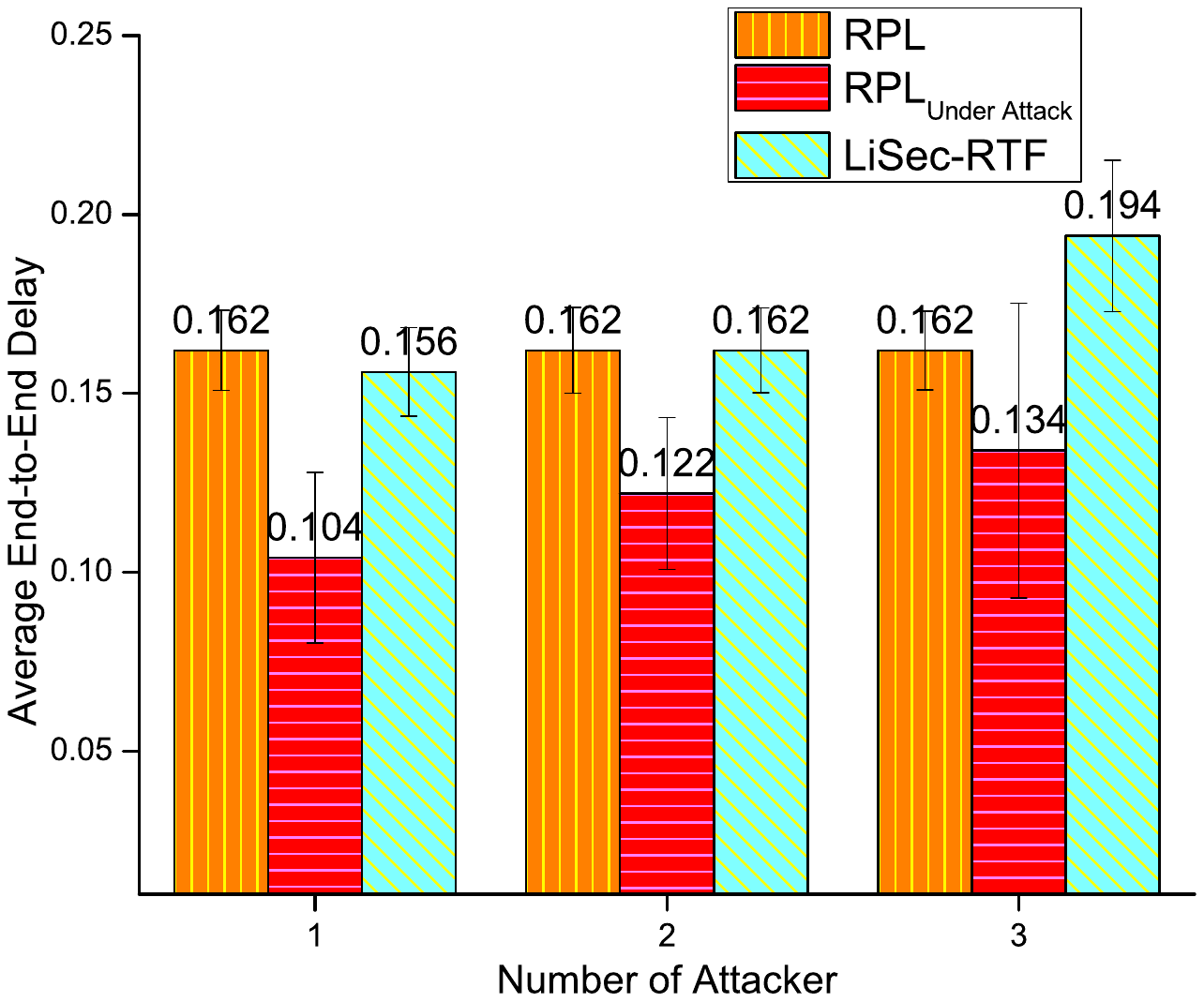}
    \caption{AE2ED in mobile scenario}
    \label{fig:s2e2e}  
 \end{subfigure}\\%   
   \begin{subfigure}[b]{.55\linewidth}
    \centering
    \includegraphics[width=\textwidth]{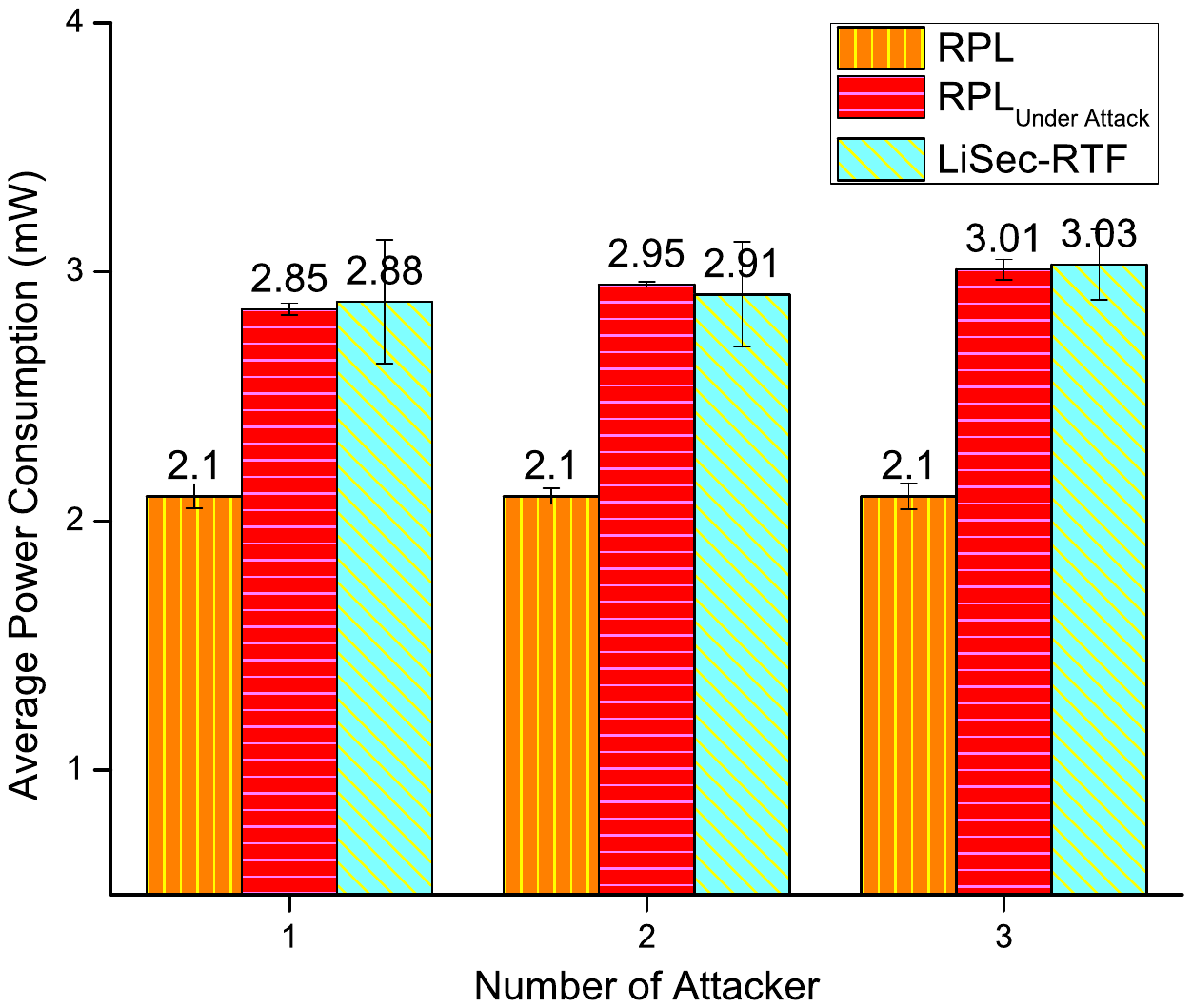}
    \caption{APC in static scenario}
    \label{fig:sapc}
\end{subfigure}%   
  \begin{subfigure}[b]{.55\linewidth}
    \centering
    \includegraphics[width=\textwidth]{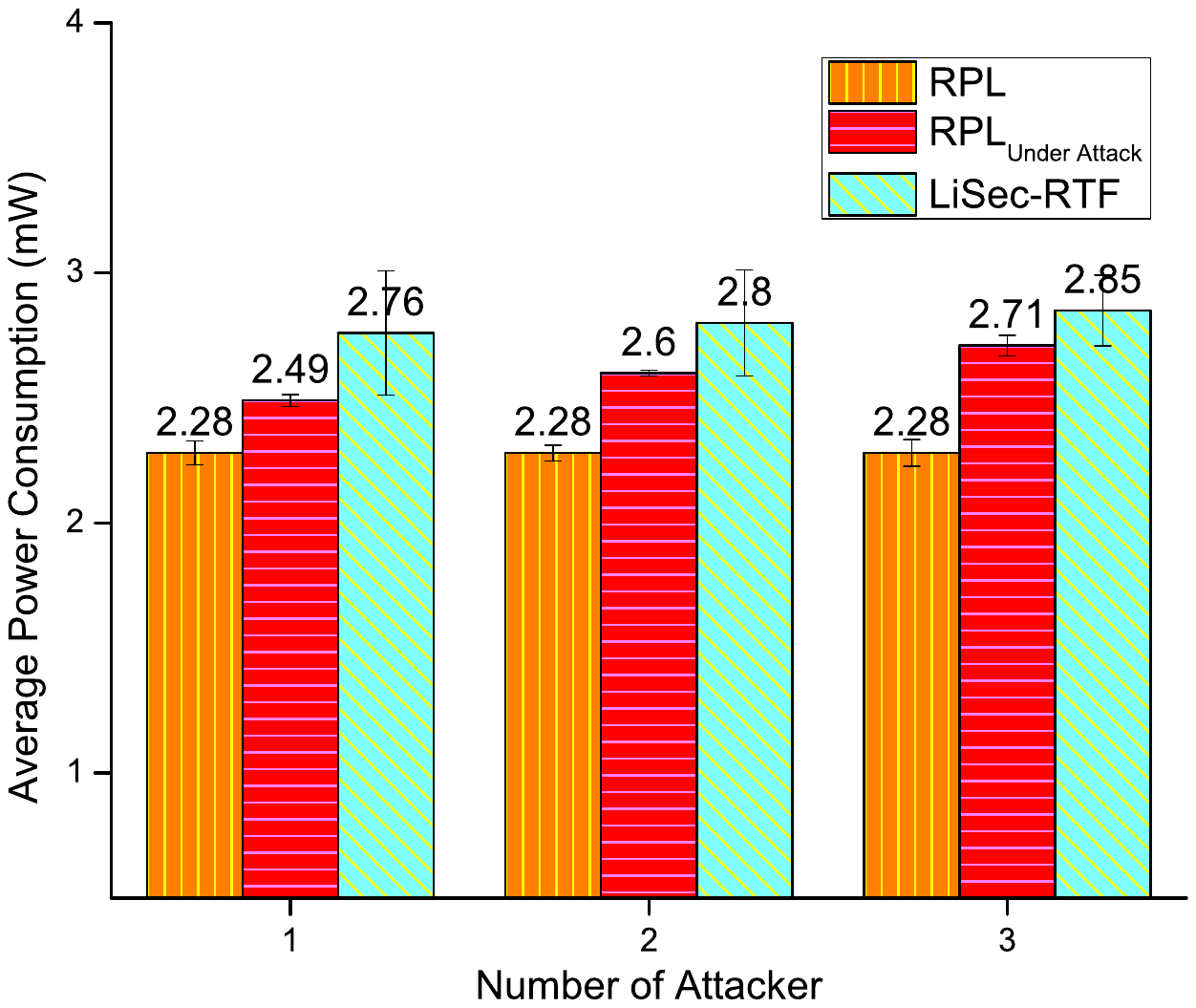}
    \caption{APC in mobile scenario}
    \label{fig:s2apc}
 \end{subfigure}\\%  
     \caption{PDR, AE2ED and APC obtained in different scenarios}
% \caption{AE2ED obtained in different scenarios}
 \end{figure}

We calculate the AE2ED by using Eq. \ref{eq8}. Figures \ref{fig:se2e} and \ref{fig:s2e2e} show the effect of AE2ED on $RPL$, $RPL_{Under~Attack}$, and LiSec-RTF in the static and mobile environments while varying the number of malicious insider attackers. In the static scenario, the AE2ED for standard RPL ($RPL$) is approximately $0.326$, while in a mobile environment, it decreases to around $0.162$. The delay of the mobile scenario is less compared to static due to the proximity of the sensor to others, and sensor nodes can dynamically change their positions and network topology. The mean range of AE2ED in $RPL_{Under~Attack}$ is $0.344$ to $0.366$ in a static environment and $0.104$ to $0.134$ in a mobile scenario. The delay in the static scenario is more than mobile because the sensor nodes may choose suboptimal or even nonexistent paths for forwarding packets. The mean values of AE2ED in the case of LiSec-RTF are $0.298$ to $0.314$ in static and $0.156$ to $0.194$ in mobile environments. The delay of LiSec-RTF is approximately the same as the $RPL$. Therefore, our proposed solution successfully addresses the RTF attack without imposing a delay on the network.

 \subsubsection{Analysis on APC}

 The RPL protocol is predominantly utilized in LLNs due to its energy-efficient routing capabilities. Hence, it is crucial to analyze the power consumption of nodes prior to deploying any new defense mechanism. As we compute the APC using Eq. \ref{eq12}, Figures \ref{fig:sapc} and \ref{fig:s2apc} illustrate the comparison of the APC on standard RPL ($RPL$), $RPL_{Under~Attack}$, and LiSec-RTF in static and mobile scenarios with varying the number of malicious nodes. The average power consumed in standard RPL ($RPL$) is $2.11$mW in static and $2.28$mW in mobile environments. The APC is more suitable for mobile scenarios compared to static because the continuous movement of the sensor nodes and the active communication between sensor nodes drain the battery faster in a mobility environment. The mean range of APC in $RPL_{Under~Attack}$ is $2.85$ to $3.01$ in static and $2.49$ to $2.71$ in mobile scenarios with varied malicious nodes. The power consumption of sensor nodes increases in static and mobile environments because, during the RTF attack, sensor nodes may need to exchange more control messages to update and synchronize their routing tables. The increased control traffic results in higher power consumption. The mean range of LiSec-RTF is $2.88$ to $3.03$ in static and $2.76$ to $2.85$ in mobility environments. Our proposed defense mechanism, LiSec-RTF, slightly increases power consumption because the nodes need to recalculate their routes and update the routing table. Therefore, the processing overhead slightly increases the power on the network.

\subsubsection{Memory Overhead}

Table \ref{tab:overhead} illustrates the memory requirements of both standard RPL and LiSec-RTF. It is generally discouraged to employ a resource-consuming defense approach within the RPL protocol. Therefore, lightweight defense mechanisms have been proposed to facilitate the creation of resource-efficient networks. Using the msp-430 size tool, this study examines the effects of integrating LiSec-RTF on RAM and ROM usage. Table \ref{tab:overhead} illustrates the memory requirements of $Mote_{RPL}$ (Contiki-NG firmware with RPL implemented) and $Mote_{LiSec-RTF}$ (Contiki-NG firmware with the proposed solution implemented). Based on the findings, there has been less than 1\% increase in the RAM and ROM requirements for LiSec-RTF. It is important to remember that the $92KB$ is the maximum storage of the standard Z1 Mote. Thus, without causing significant overhead, LiSec-RTF is appropriate for z1 motes.

 \begin{table}[!h]
	\centering
	% 	\footnotesize
	\caption{Memory Requirements}
	\label{overhead}
	\begin{tabular}{p{3cm}p{2.1cm}p{1.6cm}}\\
		\hline
		\textbf{File} & \textbf{RAM (Bytes)}& \textbf{ROM (Bytes)}\\ \hline
   udp-client.z1 ($ RPL$)  &  $6610$ & $71653$\\
   udp-client.z1 (LiSec-RTF)  &  $6610$($\mathbf{+762}$) & $71653$ ($\mathbf{+290}$)\\ \hline
\end{tabular}

    \label{tab:overhead}
\end{table}
\begin{figure}
    \centering
    \includegraphics[width=.5\textwidth]{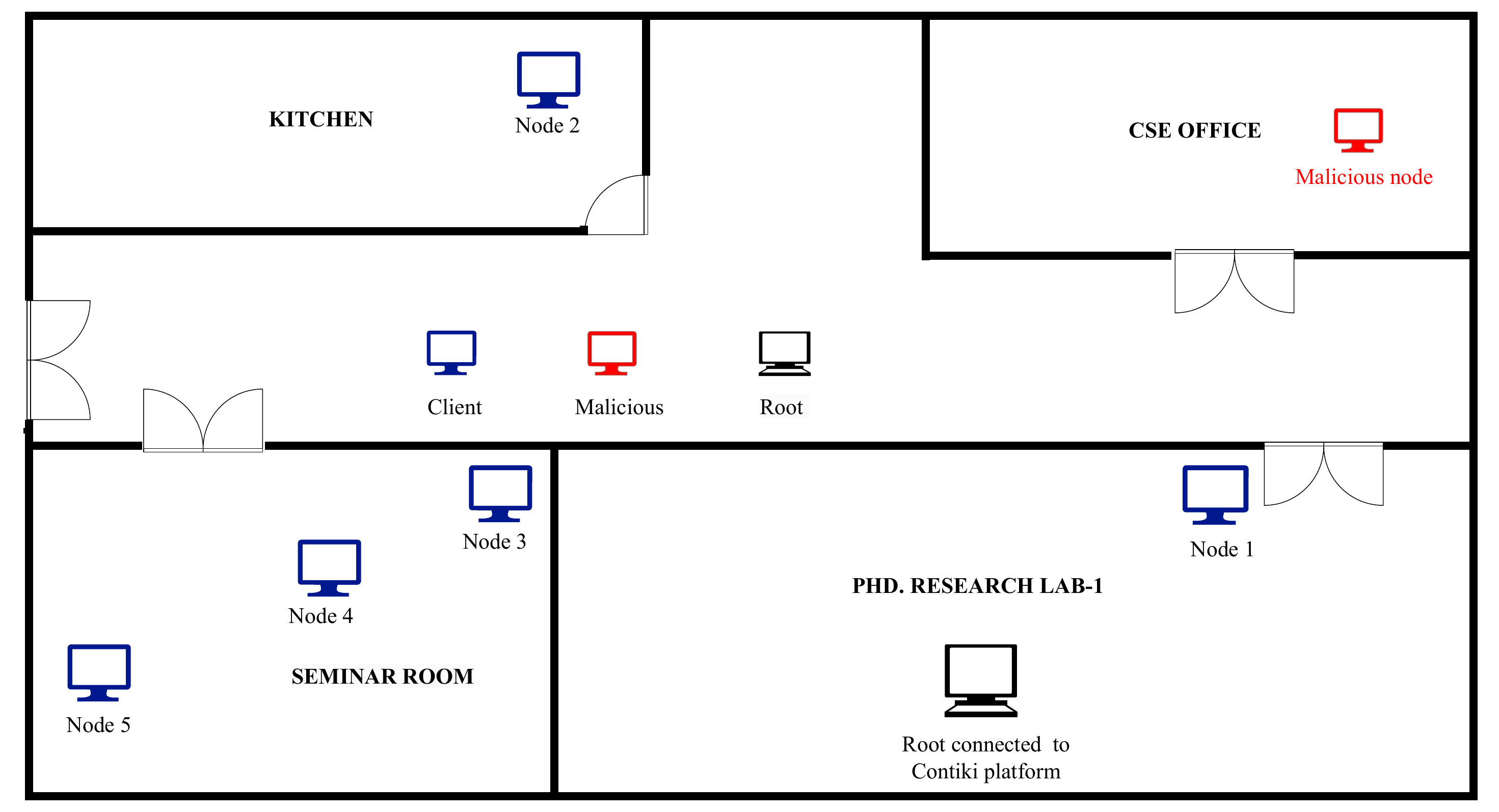}
    \caption{\textcolor{black}{A topological view of CC26X2R1 mote deployment consisting of $1$ root attached to a desktop, $5$ client nodes and $1$ malicious node.}}
    \label{fig:testbed}
\end{figure}
\subsection{Experimental Evaluation on Testbed}
\begin{figure}
    \centering
    \includegraphics[width=.5\textwidth]{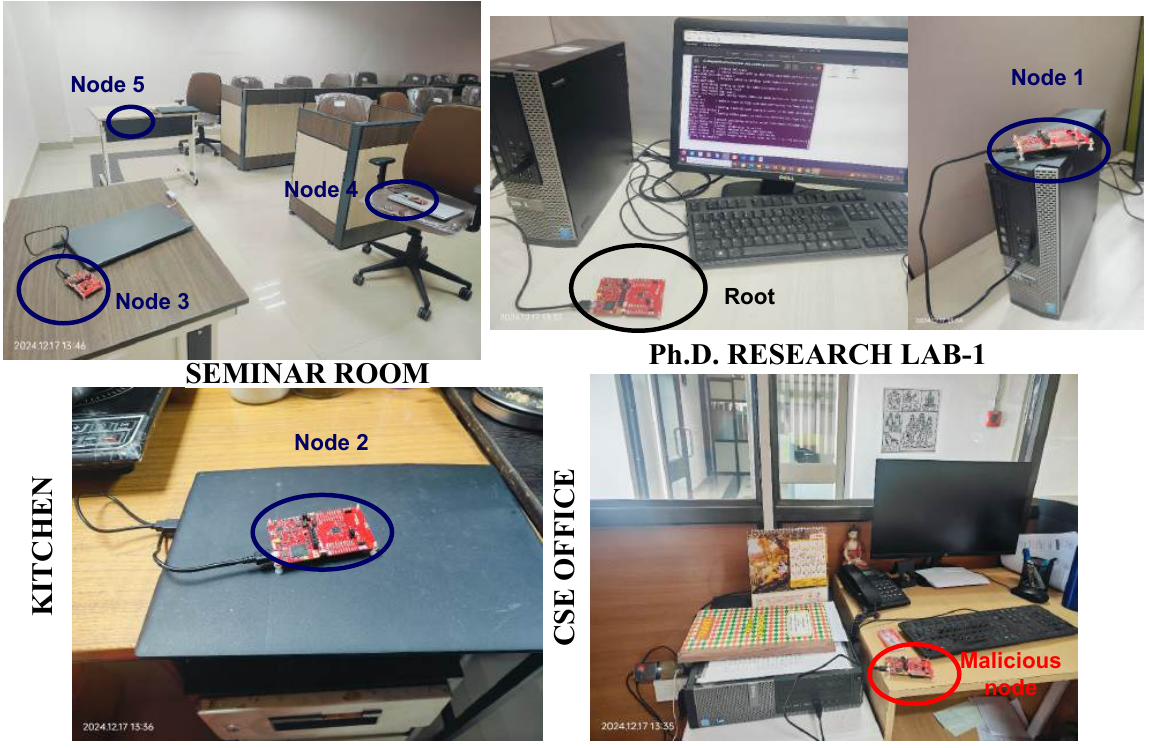}
    \caption{\textcolor{black}{Experimental testbed}}
    \label{fig:testbedexp}
\end{figure}

\begin{figure}[h!]
\centering
 \begin{subfigure}[b]{.55\linewidth}
    \centering
    \includegraphics[width=\textwidth]{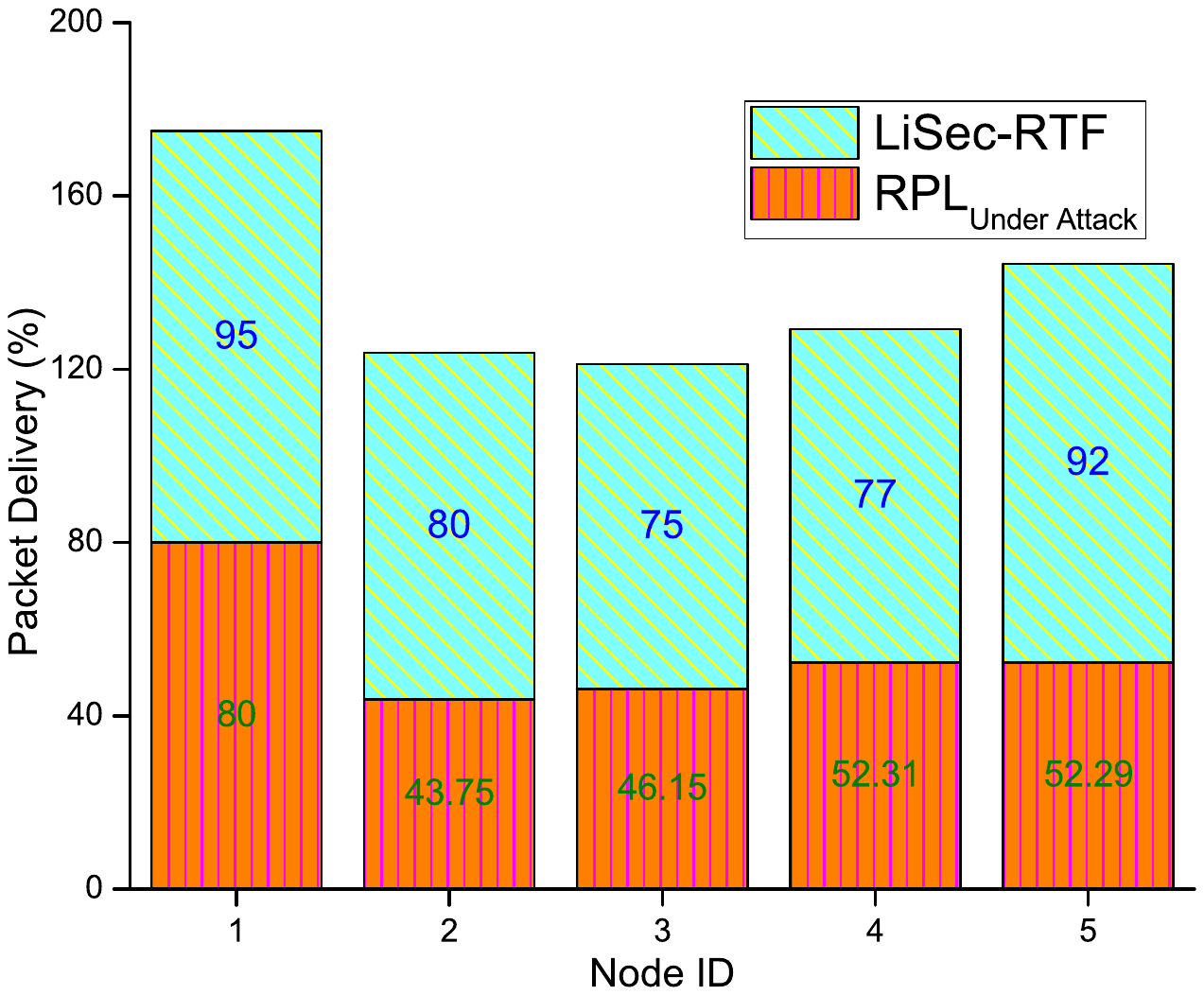}
    \caption{Packet Delivery Ratio}
    \label{fig:pdrt}
\end{subfigure}%   
  \begin{subfigure}[b]{.55\linewidth}
    \centering
    \includegraphics[width=\textwidth]{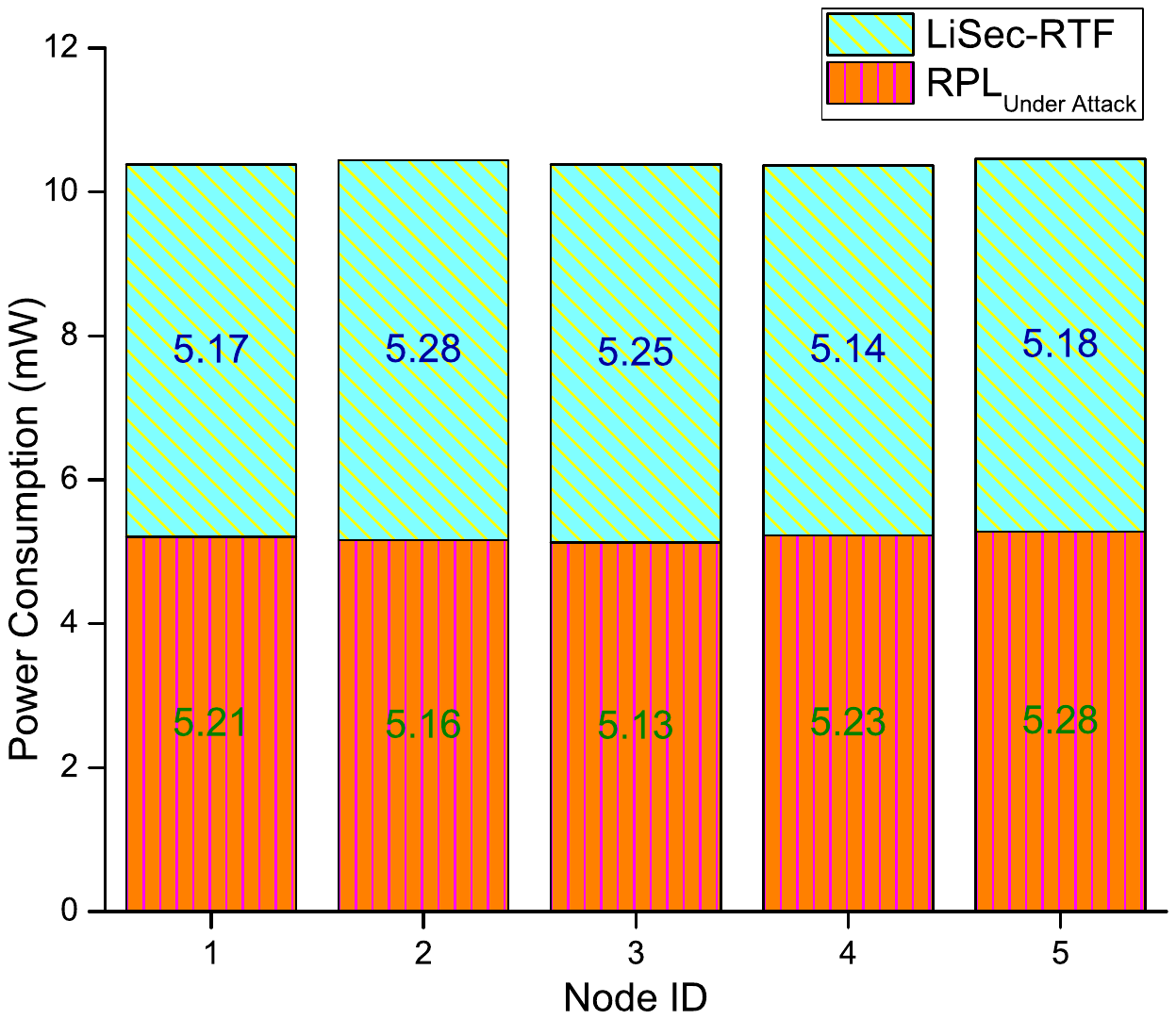}
    \caption{Power Consumption}
    \label{fig:apdrt}
 \end{subfigure}\\%  
 \caption{Experimental testbed results on packet delivery and power consumption by the nodes}
% \caption{AE2ED obtained in different scenarios}
 \end{figure}

To examine the effect of the RTF attack on the DODAG of RPL networks, we conducted a series of testbed experiments. The experiments were designed using the open-source Contiki-NG operating system to generate the necessary binary executable files. \textcolor{black}{These files were then flashed onto Texas Instruments (TI) CC26X2R1 LaunchPad devices \cite{cc26xr1}. In our experimental setup, one launchpad device was configured as the RPL root node, another as a malicious node, and the remaining devices were set up as legitimate nodes.} \textcolor{black}{For our testbed experiment, we utilized the PhD Research Lab, located on Level 3 of the CC building at IIITDM Jabalpur. Fig.\ref{fig:testbed} shows the topological view of the testbed in the lab, with $7$ CC26X2R1 sensor motes deployed for the experiment. Among them, $4$ were sender motes, including $1$ malicious mote, while $1$ mote served as the root, connected to a desktop PC running the Contiki-NG Cooja emulation program. The client and malicious motes were evenly distributed throughout the 3rd floor of the CC building. Fig. \ref{fig:testbedexp} illustrates the experimental setup of the testbed on the third floor of the CC building at IIITDM Jabalpur. The root mote was connected to the Contiki-NG platform on the PC, and $1$ additional node was deployed in the PhD Research Lab on the third floor. Additionally, the malicious mote was placed in the CSE office, while $3$ client motes were deployed in the seminar room, and $1$ mote was positioned in the kitchen area.} We executed each set of experiments five times and used the mean values of the obtained results. The root initiates the network formation, and the malicious node joins the network as a legitimate node. Figure \ref{fig:pdrt} shows the network's successful transmission of data packets. As a result of the attack, the PDR decreased by 55\%. However, our proposed approach (LiSec-RTF) improved the PDR by up to 40\% by successfully identifying the fake unicast DAO packets sent by malicious insider nodes.

Figure \ref{fig:apdrt} shows the power consumption of each node throughout the experiment. We derived the node power consumption based on its radio duty cycle, using data collected from the 'energest' module of Contiki-NG. Due to RTF attack, already-joined nodes unnecessarily transmit DAO packets to overload the parent node's routing table, leading to increased control traffic and increased power consumption. LiSec-RTF integration imposes a very minor overhead to nodes in terms of power consumption due to the need for route recalculation and routing table updates.
 \begin{figure*}
    \centering
    \includegraphics[width=.8\textwidth]{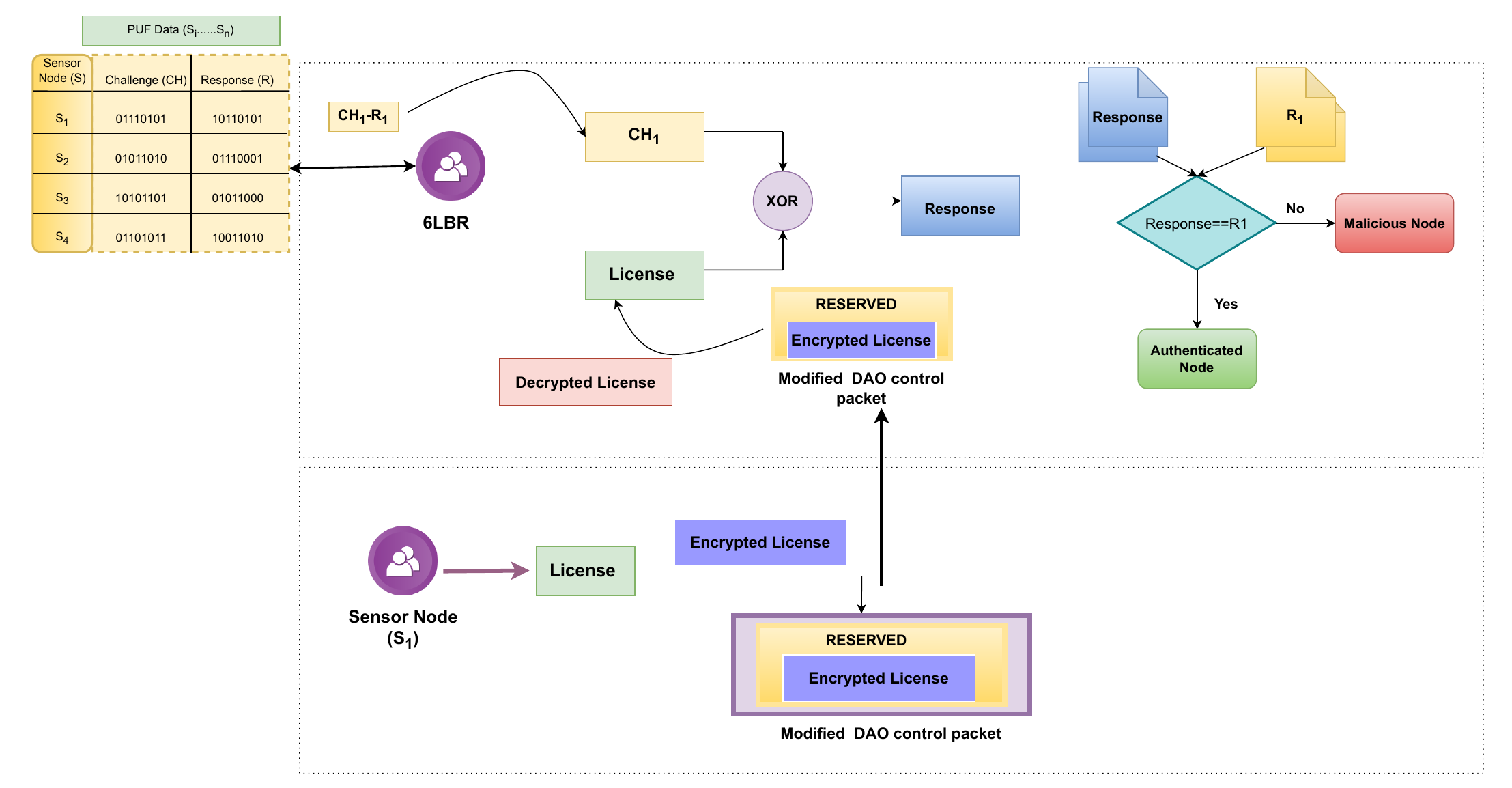}
    \caption{\textcolor{black}{Architecture of LiSec-RTF with Lightweight Encryption}}
    \label{fig:Lenc}
\end{figure*}
\section{Enhanced LiSec-RTF with Lightweight Encryption}\label{Rev3}
Several applications deal with sensitive data and have a low tolerance for security breaches. Assuming DAO messages are not encrypted, an adversary could easily obtain knowledge of the License through an eavesdropping attack. In such a case, an unencrypted LiSec-RTF would not perform well. Therefore, to address this issue, an encrypted LiSec-RTF model is also proposed. At present, various IoT hardware devices (e.g., TI CC26X2R1, CC2650, or similar low-power devices) support Advanced Encryption Standard (AES) and Elliptic Curve Cryptography (ECC) due to built-in hardware accelerators \cite{cc26xr1} Moreover, embedded operating systems like Contiki-NG and TinyOS can be easily integrated with lightweight ECC variants such as TinyECC (Tiny Elliptic Curve Cryptography), MicroECC, NanoECC, and Curve25519 \cite{rana2022lightweight, dhanda2020lightweight, szczechowiak2008nanoecc, fadia2024elliptic}. Thus, LiSec-RTF can be integrated with lightweight ECC to secure License transmission, making it difficult for an attacker to extract the exact License through eavesdropping. Consequently, this prevents unauthorized authentication at the root. The major modifications made to the unencrypted variant of LiSec-RTF are outlined below.

Instead of transmitting \(\mathcal{L}_{i}\) directly, we encrypt it using a lightweight ECC before transmission.  
The encrypted License can be represented as (Eq. \ref{eq9}):  
\begin{align}  \label{eq9}
    \mathcal{L}_{i}^{enc} = Encrypt(K_{shared}, \mathcal{L}_{i})  
\end{align}  
where \( K_{shared} \) is a pre-shared symmetric key or dynamically generated session key between the sensor node and the 6LBR.  
As shown in Fig. \ref{fig:Lenc}, upon receiving the modified DAO message, the 6LBR decrypts \(\mathcal{L}_{i}^{enc}\) before proceeding with validation.  
The equation for extracting \( r_{i} \) at the 6LBR as (Eq. \ref{eq10}):  
\begin{align}  \label{eq10}
    r_{i} = \mathcal{CH}_{i} \oplus Decrypt(K_{shared}, \mathcal{L}_{i}^{enc})  
\end{align}
Therefore by encrypting License ($\mathcal{L}_{i}$) before transmission and decrypting it at the 6LBR, we ensure that an attacker cannot extract the License from intercepted DAO's. Additionally, the use of pre-shared symmetric key ($K_{shared}$) enhances the confidentiality of the data involved in the authentication process and mitigates the risk of eavesdropping attack.

\section{Conclusion and Future Scope} \label{Sec:Conclusion}
The Routing Table Falsification (RTF) attack, an understudied routing threat initiated by an insider node, involves disseminating false information through DAO packets, leading to parent node routing table overflow. Extensive experiments reveal its adverse impact on packet delivery ratio. Given RPL's developmental stage, it lacks a defense mechanism against RTF. To address this, our paper proposes LiSec-RTF, a lightweight detection and mitigation solution. It utilizes a Physical Unclonable Function (PUF) to generate a unique authentication code (License) validated at the sink node. The widely used Cooja simulator for 6LoWPAN network analysis is used for carrying out experiments on Contiki-NG. \textcolor{black}{The experimental observations confirm the effectiveness of LiSec-RTF in mitigating RTF impact in resource-constrained static and mobile sensor node environments.} Moreover, LiSec-RTF does not impose any significant memory overhead on Zolertia z1 motes. \textcolor{black}{In this research, we have also suggested an encrypted variant of LiSec-RTF to defend against eavesdropping attacks and prevent attackers from being maliciously authenticated. Our future goal is to extend this defense solution to counteract collaborative attacks and evaluate the encrypted variant of LiSec-RTF on a testbed.}

\ifCLASSOPTIONcaptionsoff
  \newpage
\fi



%\bibliographystyle{IEEEtran}
\bibliographystyle{IEEEtran}
\end{document}